\documentclass[aps,pra,twocolumn,groupedaddress,showpacs]{revtex4}
\usepackage{color,amsmath, amssymb, graphicx, amsfonts , setspace}

\usepackage{subfigure}  

\newcommand{\dg}[0]{\ensuremath{\dagger}}
\newcommand{\ua}{\ensuremath{\uparrow}}
\newcommand{\da}{\ensuremath{\downarrow}}
\newcommand{\verysmallpic}[1]{\resizebox{3.8cm}{!}{\includegraphics{#1}}}
\newcommand{\av}[1]{\ensuremath{\left\langle #1 \right\rangle}}
\newcommand{\dif}{\ensuremath{\,\mathrm{d}}} 
\newcommand{\mi}{\ensuremath{\mathrm{i}}} 
\newcommand{\me}[1]{\ensuremath{\mathrm{e}^{#1}}} 
\newcommand{\pdo}[2]{\ensuremath{\frac{\partial{#1}}{\partial {#2}}}} 
\newcommand{\drt}[2]{\ensuremath{\frac{\dif^2{#1}}{\dif {#2}^2}}} 
\newcommand{\dro}[2]{\ensuremath{\frac{\dif{#1}}{\dif {#2}}}} 
\newcommand{\nn}{\nonumber}
\newcommand{\bra}[1]{\ensuremath{\left\langle #1\right|}}
\newcommand{\ket}[1]{\ensuremath{\left|#1\right\rangle}}

\newcommand{\matrixElement}[3]{\ensuremath{\bra{#1} #2 \ket{#3}}}

\newcommand{\abs}[1]{\ensuremath{\left| #1 \right|}}

\begin{document}


\title{Collective modes as a probe of imbalanced Fermi gases}


\author{Jonathan M. Edge}
\affiliation{T.C.M. Group, Cavendish Laboratory, J.~J.~Thomson Ave., Cambridge CB3~0HE, UK.}
\author{N. R. Cooper}
\affiliation{T.C.M. Group, Cavendish Laboratory, J.~J.~Thomson Ave., Cambridge CB3~0HE, UK.}

\affiliation{}


\date{\today}

\begin{abstract}
  We theoretically investigate the collective modes  of imbalanced two component one-dimensional Fermi gases with attractive interactions.  This is done for trapped and untrapped systems both at zero and non-zero temperature,  using self-consistent mean-field theory and the random phase approximation. We discuss how the Fulde-Ferrell-Larkin-Ovchinnikov (FFLO) state can be detected and the periodicity of the associated density modulations determined from its collective mode spectrum.
We also investigate the accuracy of the single mode approximation for low-lying collective excitations in a trap by comparing frequencies obtained via sum rules with frequencies obtained from direct collective mode calculations. It is found that, for collective excitations where the atomic clouds of the two spin species oscillate largely in phase, the single mode approximation holds well for a large parameter regime. Finally we investigate the collective mode spectrum obtained by parametric modulation of the coupling constant.

\end{abstract}


\maketitle


\section{Introduction}
\label{sec:introduction}

With the advancements in the field of ultracold atomic gases new phases of matter are becoming experimentally accessible. Superfluidity \cite{bloch07,pitaevskii_stringari_RMP_09} and density imbalance \cite{pairing_and_phase_sep_in_pol_fg_partridge_hulet,fermionic_sf_with_imbalanced_populations-zwierlein-ketterle06} in two component Fermi gases have been demonstrated experimentally. Together these offer the exciting perspective of achieving the Fulde-Ferrell-Larkin-Ovchinnikov (FFLO) phase \cite{larkin_ovchinnikov_64,superconductivity_in_spin_exch_field_fulde_ferrell64}, an unconventional superfluid state in imbalanced Fermion systems. It has been under theoretical investigation since the 1960s but has so far eluded unambiguous experimental detection. 

In solid state systems the detection of the FFLO phase is unfavourable since density imbalances between spin up and down electrons need to be imposed via Zeeman splitting, requiring  strong magnetic fields. This in turn destroys superconductivity via diamagnetic  effects \cite{landau_lifshitz_pitaevskii_stat_phys_part2}. Ultracold atomic gases are favourable for its detection since it is possible to set the density imbalance of the system simply by adjusting the number of spin up and down particles which are loaded into the atomic trap \cite{pairing_and_phase_sep_in_pol_fg_partridge_hulet,fermionic_sf_with_imbalanced_populations-zwierlein-ketterle06}. 

Theory does predict the occurrence of the FFLO phase in three-dimensional (3D) ultracold Fermi gases though the region of parameter space it occupies is thought to be relatively small (for a review see \cite{bec-bcs_xover_in_pol_res_SF-radzihovsky}). On the other hand the FFLO phase is expected to be stabilised in quasi  one-dimensional (1D) geometries which can be imposed using optical lattices \cite{bloch07,orso_attr_fermi_gases_bethe_ansatz07,phase_diag_of_str_int_pol_fg_1d_drummond07,pairing_states_of_pol_fg_feiguin_meisner07,quasi_1d_pol_fermi_SFs-parish-huse07,batrouni_scalettar_09,fflo_state_in_1d_attr_hubbard_model_luescher_laeuchli08}.

In a recent development one-dimensional imbalanced Fermi gas systems have been created in experiment \cite{spin_imbalance_in_a_1d_FG-Liao_Hulet_2009}.
These  recent experimental results indicate that the density distribution is of the form expected for the partially-polarised state associated with the FFLO phase. However, a direct demonstration of the ordering in the FFLO phase remains to be shown.
An important question which now arises, is how best to detect this characteristic ordering of the FFLO phase.  Previous proposals for the detection of the FFLO phase include  detecting characteristic peaks in the expansion image following release
of the trap  \cite{bec-bcs_xover_in_pol_res_SF-radzihovsky}, the
probing of noise correlations in an expansion image  \cite{fflo_state_in_1d_attr_hubbard_model_luescher_laeuchli08}, and the use of RF spectroscopy to excite atoms
out of the gas into other states
\cite{spectral_signatures_of_fflo_in_1d_bakhtiari_torma}.
In a recent letter \cite{signature_of_fflo_first_paper_jonathan_nigel} we  showed how the FFLO phase in 1D at $T=0$ can be detected via its collective mode signature. In this paper we extend this discussion to non-zero temperature and discuss other aspects of collective  modes in imbalanced Fermi gases.
Collective modes have proved very useful for probing the properties of ultracold balanced \cite{pitaevskii_stringari_RMP_09} and imbalanced \cite{normal_state_of_a_polarized_FG_at_unit_Lobo06,collective_osc_of_imb_FG_polaron_nascimbene_salomon,coll_ex_of_trapped_imb_FG-schaeybroeck08,trapped_2d_FG_with_pop_imbalance-schaeybroeck09} Fermi gases. Our work extends these discussions to one dimension and the presence of the FFLO state.

The paper is organised as follows. In Sec.~\ref{sec:theoretical-model} we present the theoretical model of our study. In Sec.~\ref{sec:resp-homog-syst-1} we describe the homogeneous system and present our results for the associated collective mode spectrum. We describe the speed of sound as a function of the polarisation of the Fermi gas and highlight the collective mode signature of the FFLO phase. We then discuss the effects of nonzero temperature. In Sec.~\ref{sec:coll-modes-trapp} we turn our attention to the trapped Fermi gas. We assess the accuracy of the single mode approximation for various low-frequency modes in a trap. We then investigate the effect of the FFLO phase on the collective mode spectrum of the trapped imbalanced Fermi gas. From this we deduce a signature of the presence of the FFLO phase in the collective mode spectrum. Finally we describe the response of the trapped Fermi gas to a modulation of the coupling constant.

\section{Theoretical model}
\label{sec:theoretical-model}

We study a model of a two-component Fermi gas in one dimension with attractive
contact interactions and
unequal densities.
For a homogeneous system, the exact groundstates \cite{orso_attr_fermi_gases_bethe_ansatz07,phase_diag_of_str_int_pol_fg_1d_drummond07,Phase_trans_pairing_sig_attractive_Fermi_gases-guan07,exact_anal_of_delta_fn_spin_half_attr_FG-Iida08,magnetism_and_quantum_phase_trans-he_batchelor09} and thermodynamics \cite{anal_thermodyn_and_thermometr_of_gaudin_yang_gases-zhao_oshikawa09} are known from the
Bethe ansatz
\cite{some_exact_results_for_many_body_problem_in_1d_w_delta_int-yang_prl67,un_systeme_a_une_dimension_de_fermions_en_interaction-gaudin67,thermodyn_of_1d_solvable_models-takahashi}. 
As has been discussed in detail by Liu et al. \cite{liu_drummond_07},
 Bogoliubov de Gennes (BdG) mean-field theory provides an accurate description of the exact phase
diagram for the 1D system, at least within the weak-coupling BCS
regime.
We have extended the mean-field theory to investigate the linear
response and collective modes of both the trapped and untrapped
imbalanced Fermi gas.

Our numerical calculations are based on the discretised Bogoliubov de Gennes Hamiltonian which is given by
\begin{eqnarray}
  \hat H & = &  -J\sum_{i,\sigma}\left( \hat c^\dg_{i+1, \sigma} \hat c_{i, \sigma} + h.c. \right)
  \nonumber
 \\& & 
  + \sum_i \left( \Delta_i \hat c^{\dg}_{i, \ua} \hat c^{\dg}_{i, \da} + h.c.\right)
   + \sum_{i, \sigma} W_{i, \sigma} \hat c^{\dg}_{i, \sigma} \hat c_{i, \sigma}
\label{eq:bdg}
\end{eqnarray}
where $\hat c_{i,\sigma}^{(\dg)}$ are fermionic operators for species $\sigma = \ua, \da$ on site $i$,
$W_{i, \sigma} \equiv V^{ext}_i + V^{int}_{i,\sigma} - \mu_\sigma$ (with $V^{ext}$ the external potential, $V^{int}_{i, \sigma} = U
\langle \hat c^\dag_{i,\bar\sigma} \hat c_{i,\bar\sigma}\rangle$,
$\mu_\sigma$ the chemical potentials and $\bar \sigma $ is the spin opposite to spin $\sigma$) and $\Delta_i \equiv U
\av{\hat c_{i, \da} \hat c_{i,\ua}}$ is the local superfluid gap. 
$J$ is the
hopping parameter and $U$ the on-site interaction strength ($U<0$ is
assumed throughout). 
 The angled brackets $\av{\dots}$ denote the thermal and quantum expectation value at temperature $T$.
The corresponding Bogoliubov de Gennes  equations are solved self consistently to yield the mean field  ground state.

The results we present are at sufficiently low
particle density to be representative of the continuum limit.  In the
mapping to the continuum, we relate site $i$ to position $x$ via
$x=ia$, the mass is $m = \frac{\hbar^2}{2 J a^2}$ and  the 1D interaction parameter is $g_{1d} =  Ua$. 
The interaction strength $\gamma$, defined as the ratio of the interaction energy density to the kinetic energy density, is given by
\begin{align}
  \gamma = -\frac{m g_{1d}}{\hbar^2\rho}
\end{align}
\cite{liu_drummond_07} where  $\rho$ is the total density of particles.
In the case of a trapped system we take $\gamma$ to be defined  by the density at the centre of the trap.
We restrict our study to values of $\gamma$ up to about $\gamma=1.6$ where mean-field theory gives qualitatively accurate results \cite{liu_drummond_07}. The Fermi wavevector for spin $\sigma$ is given by $k_{F\sigma} = \pi \rho_\sigma$ where $\rho_\sigma$ is the density for spin $\sigma$. We define the polarisation  $p $ of the 1D Fermi gas as
\begin{align}
  p &\equiv \frac{k_{F,\ua} - k_{F,\da}}{k_{F,\ua} + k_{F,\da}} \,.
\end{align}
Additionally, we shall find it useful to define  $ k_B T_F =E_F \equiv \pi^2\rho^2 J/4$.
In the remainder of this paper we set $\hbar=1$.

The collective mode spectrum is obtained by considering the linear response of a system to external perturbations. Divergences in the
response appear at the frequencies of the collective modes.
We obtain the linear response of the equilibrium system  by
supplementing
the Bogoliubov de Gennes equations with a self consistent random phase
approximation (RPA).
As was pointed out in \cite{signature_of_fflo_first_paper_jonathan_nigel} and as we will argue in Sec.~\ref{sec:cons-mean-field}, this approximate theory correctly captures all qualitative features of the collective modes that are important for our purposes.

The self consistent linear response within RPA is computed as follows.
Let
\begin{align}
  \hat A_{i}(t) &=
  \left(
    \hat\rho_{i,\ua}(t), \hat\kappa^\dg_i(t), \hat\kappa_i(r), \hat\rho_{i,\da} (t)
  \right)
\end{align}
where $\hat \rho_{i, \sigma} = \hat c^\dg_{i,\sigma} (t)\hat c_{i,\sigma}(t)$, $\hat\kappa^\dg_i(t) = \hat c^\dg_{i,\ua} (t)\hat c^\dg_{i,\da}(t)$ in the Heisenberg representation. Furthermore, let $\hat A_{i, \alpha}(t)$ ($\alpha \in \{1..4\}$) refer to the four components of $\hat A_{i}(t)$ and denote
  the Fourier transforms of $\hat A_{i, \alpha}(t)$ as $\hat{A}_{i,\alpha}(\omega)$.
The bare response or single particle response of $\hat A_{i, \alpha}$ to a perturbation $\delta\hat H(\omega) = \sum_{j, \beta} \delta h_{j,\beta} (\omega)\hat A_{j,\beta}(\omega)$  is given by
\begin{align}
  \av{\delta \hat A_{i,\alpha} (\omega) }&= \sum_{j,\beta} \chi^0_{i\alpha,j\beta}( \omega) \delta h_{j,\beta} (\omega)
\end{align}
where the bare response function $\chi^0$ is given by \cite{bruun_mottelson01}
\begin{align}
  \chi^0_{i\alpha,j\beta} ( \omega)
  &= \frac1\mi \int_{-\infty}^\infty \dif t \int_{-\infty}^t \dif \tau
  \av{
    \left[
      \hat A_{i\alpha}( t), \hat A_{j\beta}(\tau)
    \right]
  }
  \me{\mi\omega t}
  \label{eq_for_chi_0}
\end{align}
The thermal average $\av{\dots}$ is computed by transforming to the quasiparticle eigenbasis in which $\av{\alpha^\dagger_\mu \alpha_\nu} = \delta_{\mu,\nu} f(E_\mu)$ where $\alpha^\dagger_\mu$ is the quasiparticle creation operator for a quasiparticle with energy $E_\mu$ and $f$ is the Fermi function.
The self consistent response function $\chi$ is obtained within RPA by taking  into account the change in the Hamiltonian due to a change in the quantities $\av{\hat A_{i,\alpha}}$.
The perturbing potential will then consist of the external perturbation plus the potential arising from the interaction,
\begin{align}
  \delta \hat H &=
  \sum_{j,\beta}
  \left(
     \delta h^{\text{ext}}_{j,\beta} \hat A_{j,\beta}(\omega)
  + \sum_{k,\gamma} H^{\text{int}}_{j\beta, k\gamma} \hat A_{j,\beta} \av{\delta\hat A_{k,\gamma}}
  \right)
\end{align}
where \cite{bruun_mottelson01}
\begin{align}
  H^{\text{int}}_{l\delta,k\gamma} &= U \delta_{l,k} K_{\delta,\gamma} &
  \text{and} \qquad K_{\delta,\gamma}&=
  \begin{pmatrix}
    0 & 0 & 0 & 1\\
    0 & 0&1&0\\
    0 & 1 & 0 & 0\\
    1 & 0 & 0 & 0
  \end{pmatrix} \,.
  \label{def_of_K}
\end{align}
Inserting this into the bare response we obtain:
\begin{align}
  \av{\delta \hat A_{k,\gamma} (\omega) } &=
  \left(
    \delta_{i,k}\delta_{\alpha,\gamma} - \chi^0_{i\alpha,l\delta} H^{\text{int}}_{l\delta,k\gamma}
  \right) ^{-1} \chi^0_{i\alpha,j\beta}\delta h^{\text{ext}}_{j,\beta}
  \label{self_cons_response}
\end{align}
(summation implied), which determines the RPA result for the susceptibility.

\section{Homogeneous system}
\label{sec:resp-homog-syst-1}

\begin{figure}[tb]
  \centering
  \includegraphics[width=8cm]{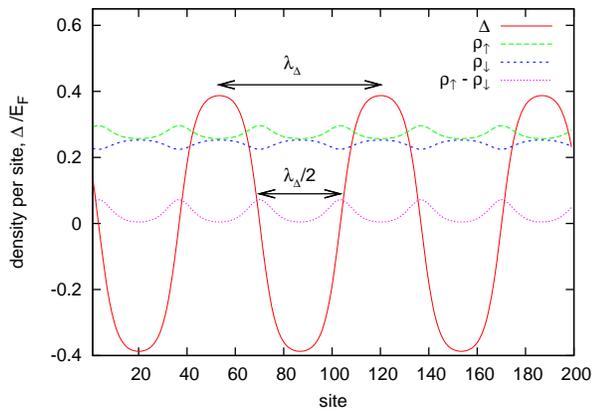}
  \caption{(Colour online)  Example of a homogeneous system displaying an FFLO
    phase. The periodicity of the densities is given by
    $\lambda_\Delta /2$ where $\lambda_\Delta$ is the wavelength of
    the order parameter. This configuration has $\gamma = 1.6$ and the
  polarisation $p=0.06$.}
  \label{fig:eg_of_homog_sys}
\end{figure}

The ground state of a homogeneous Fermi gas with contact interactions can be found exactly by the Bethe ansatz \cite{orso_attr_fermi_gases_bethe_ansatz07,phase_diag_of_str_int_pol_fg_1d_drummond07}. For attractive interactions the phase diagram spanned by the two chemical potentials at $T = 0$ contains a fully paired region, a fully polarised  (and therefore non-interacting) region and a partially polarised region. The latter is associated with the FFLO phase. The mean field  ground state was shown to be in good qualitative agreement with the exact ground state \cite{liu_drummond_07}.
The FFLO phase is characterised by an oscillating order parameter with
a wavelength $\lambda_\Delta $ and different average densities $
\rho_\sigma $ for the $\ua$ and $\da$ particles.
As the pairing suppresses density differences, 
the density difference $\rho_\ua(x) - \rho_\da (x) $ is maximal at the
nodes of the order parameter and minimal at the anti-nodes. For low
polarisations it is possible to describe the system as a fully paired
superfluid on top of which there is a gas of excess majority spin
particles. These are situated at the nodes of the order parameter
\cite{Yang_inhomogeneous_sc_state_1d_01,mizushima07}.
At
zero temperature the wavevector $k_\Delta =
\frac{2\pi}{\lambda_\Delta} $ for the order parameter is given by
$k_\Delta = k_{F,\ua} - k_{F \da}$. A possible ground state
configuration where these quantities are illustrated is shown in Fig.~\ref{fig:eg_of_homog_sys}.
With increasing temperature the order parameter disappears.
For interaction strengths we are considering, the FFLO phase disappears
at temperatures $T$ on the order of $T/T_F = 0.1 $
\cite{finite_temp_phase_diag_of_spin_pol_fg-liu_drummond08}. 

\subsection{Sound velocities in an imbalanced gas}
\label{sec:sound-velocities-an}

We have investigated the speed of sound in a homogeneous system and how it depends on polarisation.  For this we compute the response of the system to potentials of the form
 \begin{equation}
\delta W_{\sigma}(x,t) = V_{\sigma} \sin kx \cos\omega t \,.
\label{eq:vlambda}
\end{equation}
We refer to the case $V_\ua = V_\da = V_0$ as a ``spin-symmetric'' and
$V_\ua = -V_\da = V_0$ as a ``spin-asymmetric'' excitation.
Eq.~(\ref{eq:vlambda}) represents a standing wave perturbation with a fixed phase. It is sufficient to consider this type of perturbation  since we find that changing the phase of the perturbation with respect to the phase of the order parameter does not change our results in any significant way.
Throughout this paper we define the response $R$ as
\begin{align}
  R&= \sqrt{\sum_i
  \left(\delta \rho_{i, \ua}^2 + \delta \rho_{i,\da}^2\right)}
\end{align}
 which is a measure of the change in the density.

\begin{figure}
  \centering
  \includegraphics[width = 8.5cm]{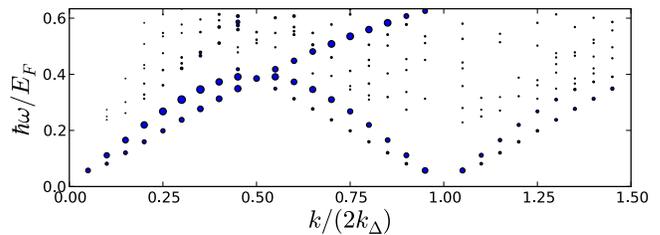}
  \caption{(Colour online)  Response
    of a homogeneous system in the FFLO phase to a
    spin-asymmetric periodic potential (\protect\ref{eq:vlambda}).
    The area of the circle is proportional to the amplitude of the response.
    The
  polarisation is $p  = 0.15$, and the
  interactions strength is $\gamma = 1.5$.  Two gapless sound modes are seen
  to emerge around $k\simeq 0$ and $k\simeq k^* \equiv 2 k_\Delta$. The simulation was done on 270 lattice sites. The divergent response  at
  $\omega=0$ is not shown.
  }
  \label{fig:two_sound_modes}
\end{figure}

\begin{figure}[tb]
  \centering
  \includegraphics[width=8cm]{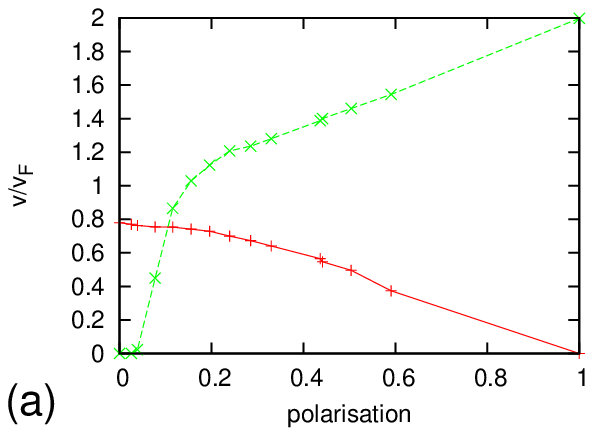}
  \includegraphics[width=8cm]{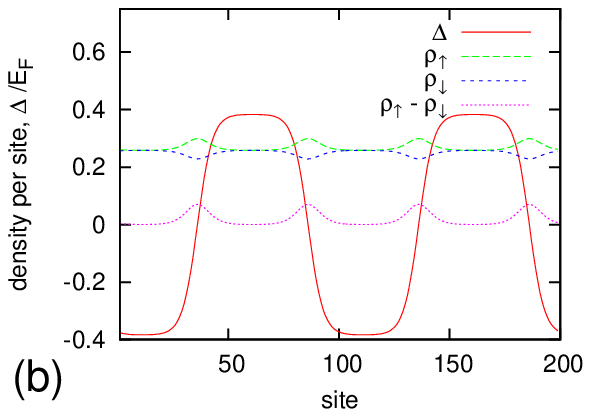}
  \caption{(Colour online) (a) Sound velocity versus polarisation for interaction
    strength $\gamma \approx 1.6$. $v_F = \frac12(v_{F,\ua} + v_{F,
      \da})$. As $p \to 0$  the velocity of one of the sound modes
    disappears. As soon as the domain walls between regions of
    positive $\Delta$ and negative $\Delta$ cease to overlap  the sound velocity drops below our numerical resolution. This is because shifting the position of a domain wall costs very little energy.  (b) Order parameter and density distributions for  a polarisation ($p=0.038$) at which the domain walls cease to overlap.}
  \label{fig:sound_vs_polarisation}
\end{figure}

For nonzero polarisations when the FFLO state is present two distinct sound modes exist with in general different velocities, as shown in Fig.~\ref{fig:two_sound_modes}. This can be understood within mean field theory from the fact that the FFLO phase breaks both gauge and translational symmetry.
However, as we will argue in Sec.~\ref{sec:cons-mean-field}, these results are fully consistent with those expected from exact theory.
The sound velocities at $T=0$  of these two modes as a function of the polarisation are shown in
Fig.~\ref{fig:sound_vs_polarisation}. (Small fluctuations in the
lattice filling due to the technique used lead to  small variations
in the sound velocity
\cite{sound_velocity_and_dim_crossover_in_sf_FG-koponen-Torma}.) For
very low polarisations we effectively have a paired superfluid on
top of which there is a weakly interacting gas of excess majority spin
 particles,  which are located at the positions of the domain walls in
 the order parameter \cite{Yang_inhomogeneous_sc_state_1d_01,mizushima07}. In this regime of low polarisation one of the sound
modes is associated with phase fluctuations of the order parameter, and has a very similar velocity as that for a spin balanced
superfluid. The other sound mode, with lower velocity, arises from  the motions of the excess majority spin particles (the domain walls). The speed of sound of this mode becomes very small (below the resolution of our calculation) at very low polarisations, as the domain walls become very dilute and cease to overlap.
In the other limit, $p\to 1$, one of the sound modes vanishes, whereas the higher velocity mode tends to the value for a non-interacting Fermi gas. 

\subsection{Signature of the FFLO phase}
\label{sec:signature-fflo-phase}

We now turn to the signature of the FFLO phase in the collective mode spectrum.
Again we investigate the response of the system to potentials of the form (\ref{eq:vlambda}). However, this time we concentrate on short wavelength perturbations.

For zero temperature there is a clear signature of the FFLO
phase in the response spectrum of a homogeneous Fermi gas. As
illustrated in Fig.~\ref{fig:two_sound_modes} this
manifests itself in sound modes emerging around $k = k^* = 2k_\Delta$
in addition to the ones emerging around $k=0$ \cite{signature_of_fflo_first_paper_jonathan_nigel}.
Within mean-field theory, this can
be understood as a consequence of the broken translational symmetry of
the FFLO phase, though this too is  fully consistent with exact theory (see Sec.~\ref{sec:cons-mean-field}).

These low energy sound modes are due to a Brillouin zone (BZ) arising from the spatial periodicity of the superfluid order parameter $\Delta(x)$.
The consequence is a
 generalised Bloch theorem.
Let $\hat T $ be the operator which performs the transformation:
$
x \to x + \lambda_\Delta/2
,\;
\Delta(x) \to -\Delta(x)
$.
For a system with an FFLO wavelength of $\lambda_\Delta $, $\hat T $ commutes with the  Hamiltonian (\ref{eq:bdg}) since $\rho_\ua$ and $\rho_\da$ have periodicities $\lambda_\Delta/2$. Note that $\hat T$ encodes the smallest translational shift for which such a symmetry exists.
As the Hamiltonian commutes with $\hat T $, the density matrix $\hat\rho_0 $ for the equilibrium configuration is invariant under $\hat T^{-1}\hat\rho_0 \hat T $ and $\hat T $ is time-independent. In the continuum limit the response of the system is given by
\begin{align}
  \av{\delta \hat A_\alpha (r,t)} &= \sum_\beta \int_{-\infty}^t\dif \tau \int \dif r' \;\chi^0_{\alpha,\beta}(r, r', t, \tau)\; \delta h_{\beta} (r',\tau)
  .
\end{align}
Within linear response $\chi^0$ is given by \cite{forster75}
\begin{align}
  \chi^0_{\alpha,\beta}(r,r',t, \tau)
  &= \frac1\mi \av
  {
    \left[
      \hat A_\alpha(r,t), \hat A_\beta(r',\tau)
    \right]
  }
  \label{def_of_chi_in_continuum}
  .
\end{align}
As described above, in the self consistent random phase approximation $\delta \hat H$ is given by an externally applied perturbation plus a self consistent perturbation. Consider free oscillations, i.e. the external perturbation is set to zero.
Then  $\delta \hat H$ is given by $\delta \hat H(\tau) =\sum_\beta \int \dif r'\; g_{1d} K_{\beta,\gamma} \av{\delta \hat A_\gamma (r',\tau)}$ with $K_{\beta,\gamma}$ given by eq.~(\ref{def_of_K}). The equations of motion are then given by 
\begin{align}
  \pdo{}t \av{\delta \hat A_\alpha (r,t)} &=  \int \dif r' \;\chi^0_{\alpha,\beta}(r, r', t, t) g_{1d} K_{\beta,\gamma} \av{\delta \hat A_\gamma (r',t)}
  \nn
  \\&
  +\int_{-\infty}^t \dif \tau  \int\dif r' \;\pdo{}t \chi^0_{\alpha,\beta}(r, r', t, \tau)
  \nn
  \\&
  g_{1d} K_{\beta,\gamma} \av{\delta \hat A_\gamma (r',\tau)}
\end{align}
(summation implied).
The equations of motion are invariant under the operator $\hat T $  provided $\chi^0$ is invariant under $\hat T $. This can be easily  shown to be the case using the fact that $\hat T^{-1}\hat\rho_0 \hat T = \hat\rho_0$ together with eq.~(\ref{def_of_chi_in_continuum}).

\begin{figure*}[t]
  \centerline{
    \verysmallpic{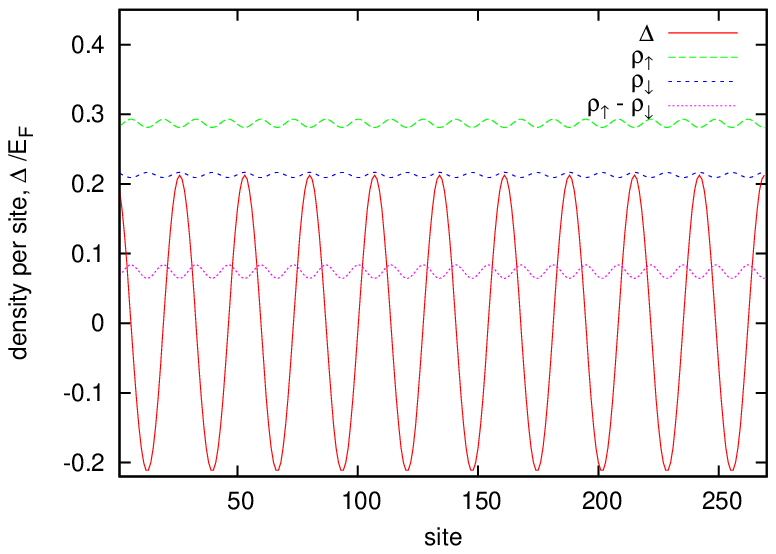}
    \verysmallpic{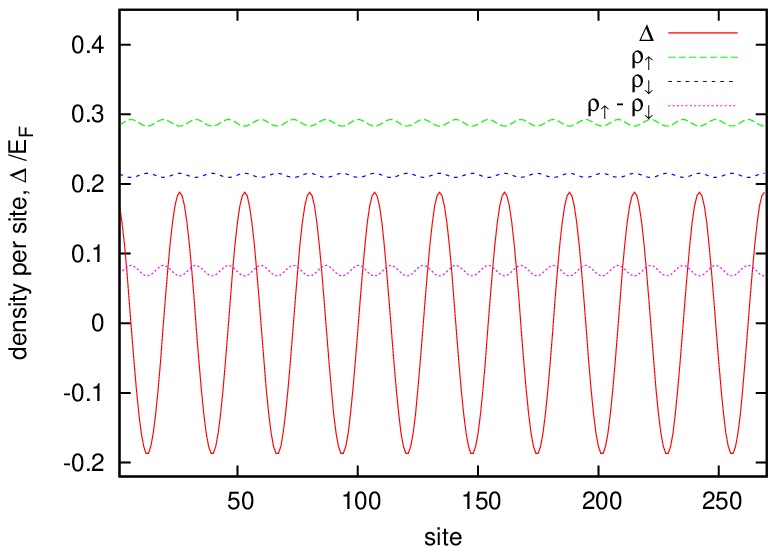}
    \verysmallpic{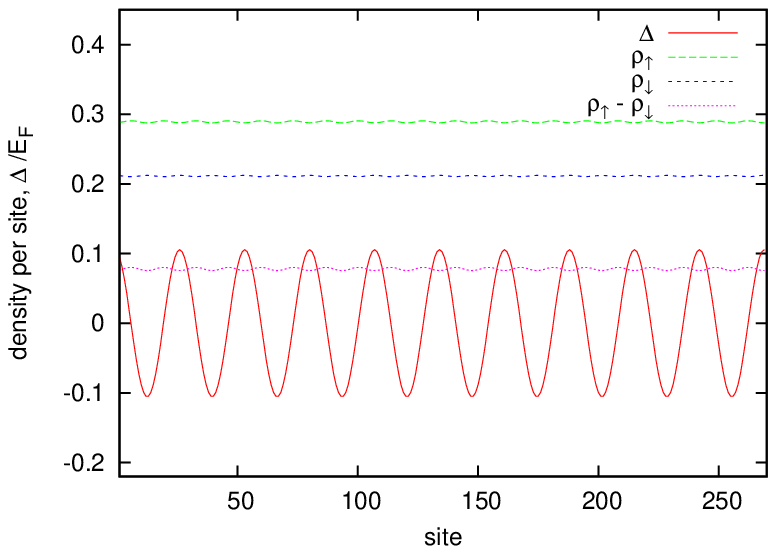}
    \verysmallpic{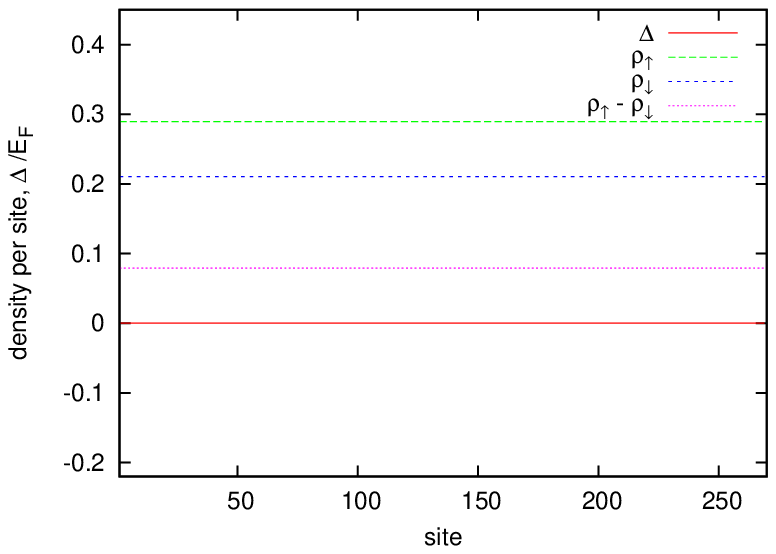} 
  }

  \centerline{
    \verysmallpic{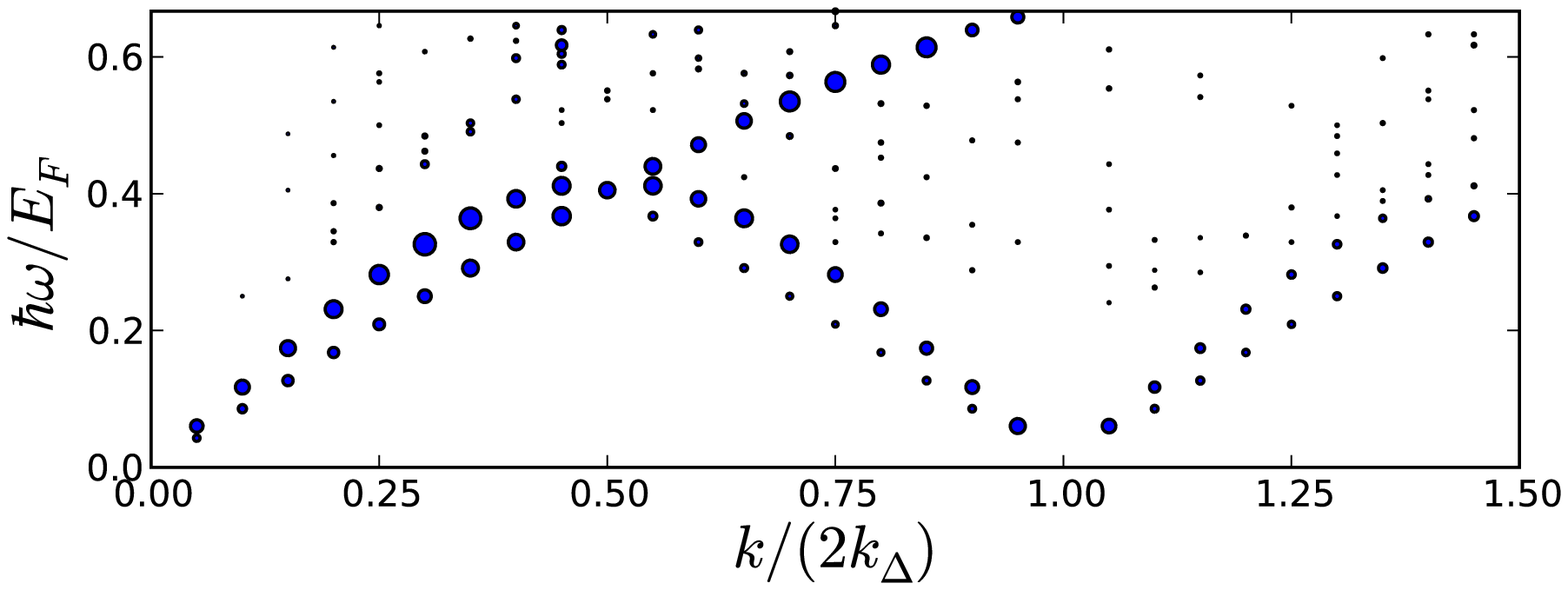}
    \verysmallpic{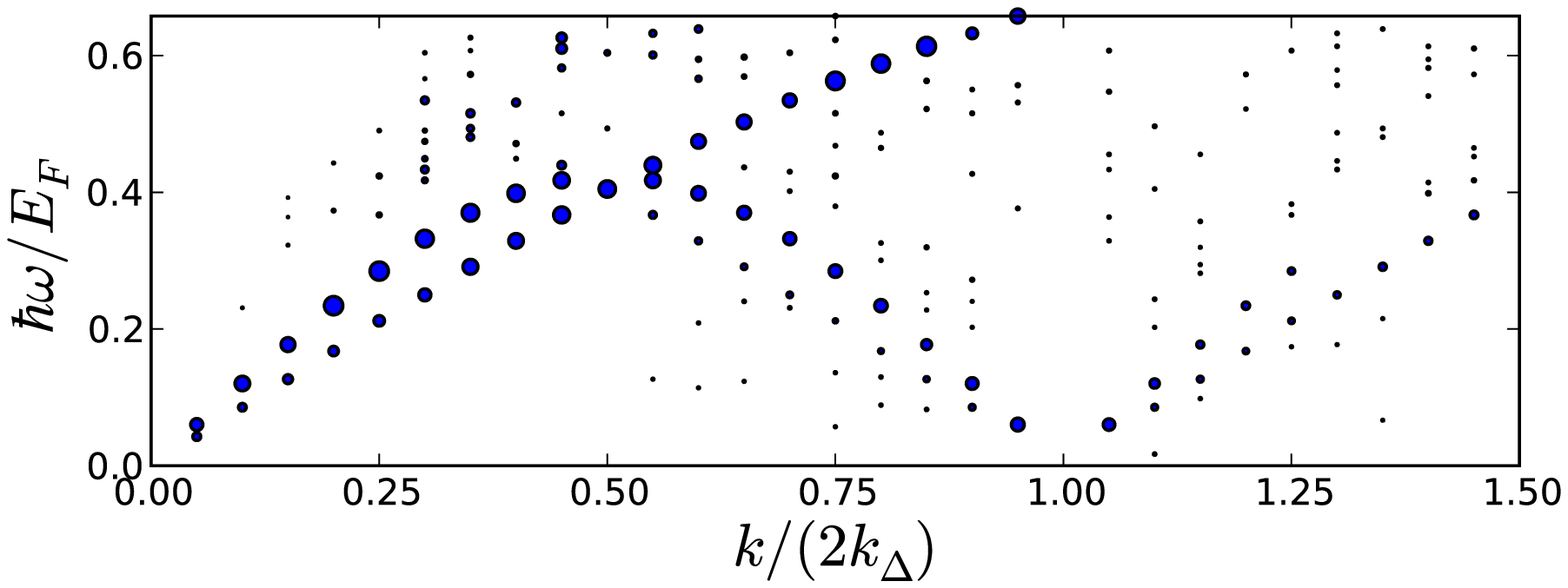}
    \verysmallpic{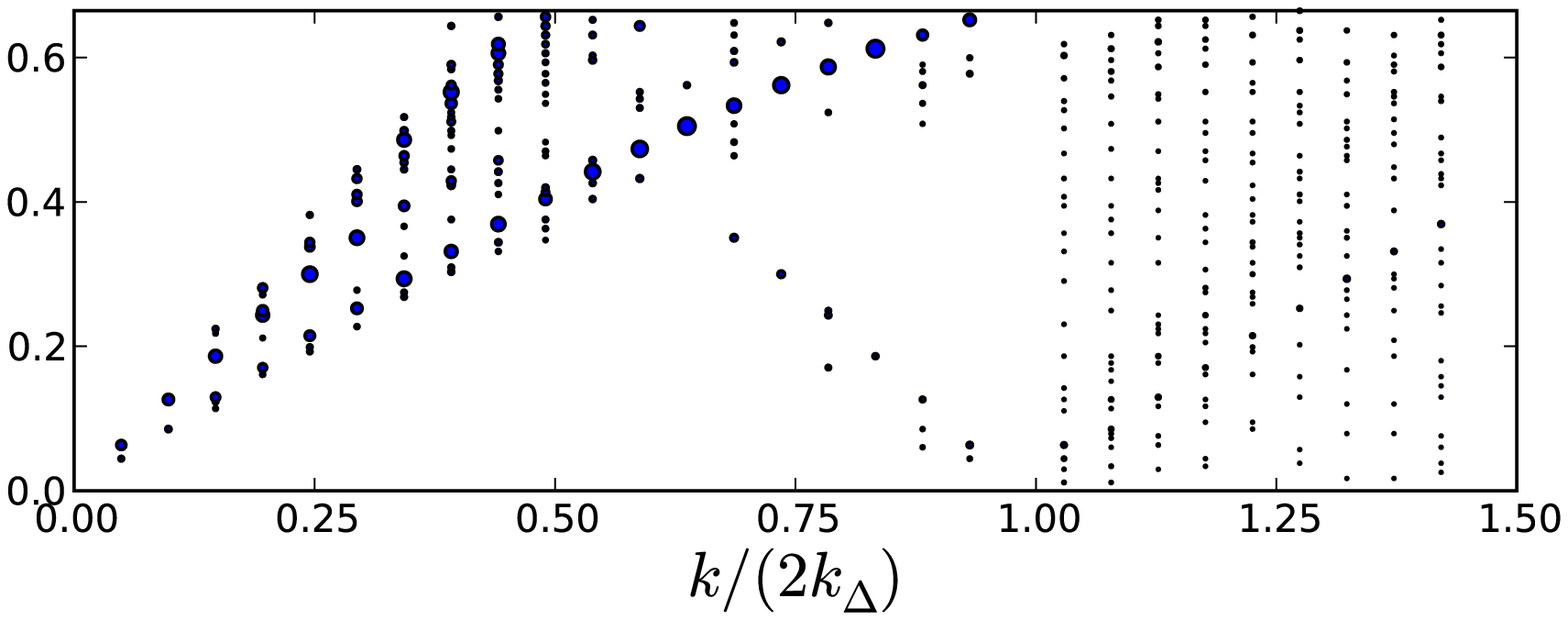}
    \verysmallpic{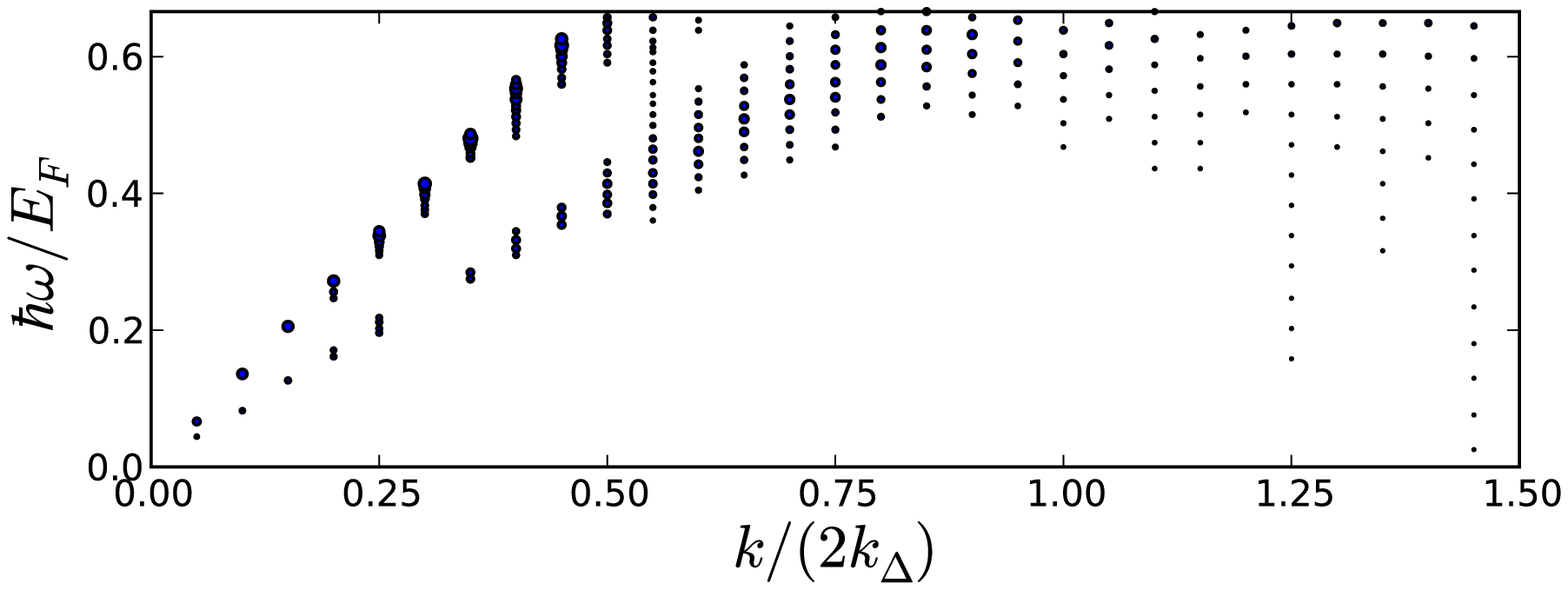}
  }
  
  \caption{(Colour online)  Response of a system at fixed chemical potential and different temperatures. The area of the circle is proportional to the amplitude of the response, as in Fig.~\ref{fig:two_sound_modes}.
    The
  polarisation is $p  = 0.15$, the
  interactions strength is $\gamma = 1.5$ and there are roughly 135 particles. 
    The temperatures are from left to right $T/T_F = 0.0037, 0.038, 0.057, 0.076$. Within our mean field theory calculation the oscillating order parameter disappears at a temperature $T/T_F = 0.07$. Simultaneous with the disappearing of the oscillating order parameter the response around $k=k^*$ disappears.
  }
  \label{fig:finite_temp2}
\end{figure*}

The result of this generalised Bloch theorem is that $k^* \equiv
(2\pi)/(\lambda_\Delta/2) = 2k_\Delta$ is a reciprocal lattice vector
for the collective mode spectrum.  This means that the conserved
  quasimomentum is only defined up to multiples of $k^*$.
Thus it is possible to perturb the system at a wavevector $q $ and obtain a response at a wavevector $q - nk^* $ ($n $ integer) which allows low-frequency sound modes to be excited when $q \approx n k^*$. Note that the generalised Bloch Theorem only states that this is possible. The strength of this response needs to be calculated separately and is determined by the matrix element coupling perturbation $q $ to response at $q - nk^* $. A smaller amplitude of the oscillating densities, e.g. due to higher temperature, will lead to a weaker coupling of these responses, see fig.~\ref{fig:finite_temp2}. Once the system becomes homogeneous the Brillouin zone structure disappears and the coupling vanishes.
Note that this argument implies a BZ twice the size than if the periodicity was simply given by $\lambda_\Delta$. %

\subsection{Mean field theory versus exact theory}
\label{sec:cons-mean-field}

While the calculations we have presented are within RPA mean-field
theory, as we now argue, the qualitative conclusions are valid more
generally.
In the above, the response spectrum was accounted for in terms of the
broken translational and gauge symmetry of the mean-field state. As is well
known, in a true 1D quantum system no continuous symmetries are
broken. Thus, the broken (phase and translational) symmetries of the
mean-field FFLO state can lead only to power-law decay of the
respective correlation
functions \cite{fflo_pairing_in_1d_opt_latt-rizzi_fazio08}.  However,
the qualitative features described above are the same as those
expected from an exact treatment of the system. The transition from the
(unpolarised) superfluid phase to the partially polarised phase
is marked by the closing of the spin-gap, leading to a second gapless
sound mode. This can be viewed as a Luttinger liquid representing the
excess fermions \cite{Yang_inhomogeneous_sc_state_1d_01}.
  These excess
particles have density $\rho_\ua-\rho_\da$, and thus a Fermi wavevector
$(k_{F,\ua} - k_{F,\da})$.  Thus, there are two gapless collective
modes, arising from the fully paired particles, and the liquid of
excess majority spin particles.  Furthermore, as in the general theory
of Luttinger liquids \cite{HaldaneJPhysC} the spectral function of
these excess majority spin particles will show gapless response at
multiples of twice their Fermi wavevector. This is $2(k_{F,\ua} -
k_{F,\da}) = 4\pi/\lambda_\Delta$ which is precisely the wavevector
$k^*$ at which one finds the gapless response in mean field
theory. 
 Thus, the qualitative features of the collective
excitation spectrum obtained in mean-field theory are fully consistent
with those expected for the exact system.

\subsection{Temperature dependence of the collective modes}
\label{sec:temp-depend}

We now consider the temperature dependence of the collective response of the homogeneous system.
For very low temperature ($T \lesssim 0.01T_F $) we find essentially the
same response spectrum as for zero temperature. As the temperature of the system is
increased and the order parameter disappears, the low energy response
around $k = k^* $ also disappears, see Fig.~\ref{fig:finite_temp2}. This can be explained by the fact that the BZ associated with the oscillating order parameter disappears.
We also observe that as the temperature is increased $k^*$ no longer coincides with $2(k_{F,\ua} - k_{F,\da})$. We note that thermally excited quasiparticles which do not contribute to the pairing can lead to this departure.

The results we have presented so far are purely within the
collisionless regime. The collisionless regime describes the situation
where the rate of quasiparticle collisions is much smaller than the
frequency of the excitation. At  zero temperature the system is always
in the collisionless regime. On the other hand for non-zero
temperature there are thermally excited quasiparticles. These may
undergo collisions, thereby leading to a nonzero collision rate. As
pointed out in Ref. \cite{theory_of_quantum_liquids1_nozieres_1966}, at nonzero temperature great care has to be taken when taking the limits $k \to 0 $, $T \to 0 $. These two limits do not commute. If first the temperature is taken to zero and subsequently $k $ is taken to zero, the system is in the collisionless regime. Sound modes in the collisionless regime correspond to zero sound. On the other hand if first $k \to 0$ and subsequently $T\to 0$ a hydrodynamic description is needed. In the hydrodynamic regime quasiparticles have time to scatter many times during the cycle of one collective oscillation and hence the system can be described in terms of local thermodynamic equilibria. In the crossover region no sound propagation is possible due to strong damping.

We have performed a simple analysis of the quasiparticle collision rate. In an infinite one-dimensional system no quasiparticle excitations can exist due to the divergent nature of the interactions; a Luttinger liquid \cite{HaldaneJPhysC} description of the system is required \cite{quantum_physics_in_1d_giamarchi}. All excitations are collective excitations. This manifests itself in an infinite quasiparticle scattering rate. On the other hand, in a finite system,  the quasiparticle scattering rate is finite.

Using perturbation theory we find that  in a system of size $L$, for $T\ll T_F$, the  scattering rate of  a quasiparticle with energy $k_BT$ is
\begin{align}
  \Gamma &= \frac{g_{1d}^2 k_F^2}{8\pi\hbar E_F} \frac{\me{}}{\me{}+1}
  \left[
    \frac T{T_F} + 2 \ln
    \left(
      \frac{\sqrt 3}{4\pi} \frac T{T_F} k_F L
    \right)
  \right]\,.
  \label{eq_for_Gamma}
\end{align}

In general for an infinite system the collisionless approximation will fail since $\Gamma$ diverges. In this case a hydrodynamic theory must be used. However, in practice all experimentally relevant systems are finite. Therefore there will be some temperature below which the system can be described as collisionless.
For example,  comparing the frequencies with the quasiparticle scattering rate for the system sizes used to compute  Fig.~\ref{fig:finite_temp2}, we find that in all apart from the four lowest frequency points for $T=0.076 T_F$ and the lowest frequency point for $T=0.057T_F$ the systems are in the collisionless regime. We therefore expect the results presented in  Fig.~\ref{fig:finite_temp2} to be qualitatively accurate  for systems of this size.

For a given system of size $L$ the temperature below which the system can be considered collisionless may be found as follows. The lowest $k$ vector with which the system can be excited is given by $k=\frac{2\pi}L$. Fig.~\ref{fig:sound_vs_polarisation} suggests that for the parameters we are considering, the sound velocity $v$ is of the order of the Fermi velocity $v_F$. Then using the fact that the collisionless approximation will be valid up to approximately $\omega\approx\Gamma $ and using $\omega \approx kv_F $  we obtain that for a homogeneous system the maximum temperature $T_m$ for which the collisionless approximation is valid is given by
\begin{align}
  T_m &\approx T_F \frac{4\pi}{\sqrt 3 k_FL}
  \me{\frac{\pi^4(\me{}+1)}{\me{}} \frac1{k_FL\;\gamma^2}}
  \label{eq_for_T_m_for_collless_regime}
\end{align}
Here we we have only included the logarithmic dependence of $\Gamma $ on $T/T_F $ and not the linear part since the linear part is much smaller. Using eq.~(\ref{eq_for_T_m_for_collless_regime}) we find that for the configurations in Fig.~\ref{fig:finite_temp2} the collisionless approximation is valid up to temperatures of around $T = 0.05T_F $. For these systems  this temperature coincides roughly with the temperature for which the FFLO phase disappears.

Above we argued that for the systems and temperatures studied in this
paper we can work in the collisionless regime. The results of our
calculations are valid in the regime in which interactions are assumed
to be weak or moderately strong. On the other hand it has been
reported that for strongly interacting 3D imbalanced Fermi systems one
needs to consider the effects of collisions of quasiparticles even at
experimentally achievable low temperatures
\cite{coll_props_of_pol_FG-bruun-stringari08,collective_osc_of_imb_FG_polaron_nascimbene_salomon}.
There are some fundamental differences between those studies of 3D
systems and the 1D systems of interest here.  In the 3D systems, a
spin imbalanced phase is stable only at relatively high polarisations
(beyond the Chandrasekhar-Clogston limit). The collision rate is rapid until the
temperature is small compared to the Fermi temperature of the minority
component \cite{coll_props_of_pol_FG-bruun-stringari08}, which, in view
of the large imbalance, can be a low energy scale.  In contrast, in 1D
the FFLO phase occupies a large region of parameter space, and it is
possible to have this spin-imbalanced phase with much smaller
polarisations when the Fermi energies for spin $\ua$ and $\da$ atoms
are of the same magnitude.  One then expects that, for strong
interactions, the relevant energy scale below which collisional
scattering starts to become small is just the (typical) Fermi
temperature.  Our formulae (\ref{eq_for_Gamma}) and
(\ref{eq_for_T_m_for_collless_regime}) were derived using perturbation
theory, so cannot be applied to determine the collisional scattering
rate in the strongly interacting regime.
We are not aware of a calculation of the thermal broadening of the collective modes for a strongly interacting spin-$\frac12$ Fermi system of relevance here.

\section{Trapped system}
\label{sec:coll-modes-trapp}

We now turn our attention to the experimentally more relevant case of
the Fermi gas in a harmonic trap. 
Within the local density approximation (LDA) the possible ground state
configurations for 1D Fermi gas at $T = 0 $ are known exactly from
the Bethe ansatz
\cite{phase_diag_of_str_int_pol_fg_1d_drummond07,orso_attr_fermi_gases_bethe_ansatz07}.
There are only two possible configurations which are (a) a partially
polarised phase in the middle, with a fully paired phase towards the
edge and (b) a partially polarised phase in the middle, with a fully
polarised phase towards the edge. Recently  these configurations have also been observed in experiment \cite{spin_imbalance_in_a_1d_FG-Liao_Hulet_2009}.
 As shown by 
Ref. \cite{liu_drummond_07} Bogoliubov de Gennes theory reproduces these
ground state configurations and it is in good qualitative agreement
with the results of the exact solution combined with the LDA.

The low-frequency collective modes in a trap are the dipole and the breathing modes. Dipole  modes are the lowest lying modes with odd parity around the centre of the trap. For $V_{\text{ext}}(x) = \frac12 m\omega_0^2 x^2$ their frequencies in the non-interacting limit are given by $\omega = \omega_0$. Breathing modes are the lowest lying modes with even parity. They correspond to  expansion and contraction of the atomic cloud. In the non-interacting limit their frequencies are given by $\omega= 2\omega_0$. Both the dipole and breathing modes exist in two types: the spin and density modes. In the density modes the two atomic clouds move in phase whereas in the spin modes the two clouds move largely in anti-phase. Within these modes the density dipole mode has a special status. It is called  the ``Kohn mode'' and corresponds to rigid body oscillations of the cloud about the centre of the trap.
Its frequency $\omega $ is given by $\omega =\omega_0$ independent of the interaction strength.

We  now investigate the accuracy of the single mode approximation  for these low frequency modes.

\subsection {Single mode approximation}

The single mode approximation (SMA) is an important tool in
calculating collective mode frequencies. Together with sum rules it
allows the calculation of collective mode frequencies from ground
state properties alone. This can be done as follows: for an operator
$\hat F $ the dynamic form factor $S (\omega) $ describes the strength of the transition from the ground state to state with energy $\hbar\omega $. $S (\omega)$ is given by \cite{sum_rules_and_giant_resonances_in_nuclei-lipparini_stringari89}

\begin {align}
S (\omega)= \sum_{k\not=0} \abs{ \matrixElement {k} {\hat F} 0}^2 \delta(\omega - \omega_k)
\end {align}
where $\ket k$ label the exact energy eigenstates of the system.
We define the $p$th moment of the dynamic form factor as
\begin {align}
m_p= \int_0^\infty S(\omega) \omega^p \dif \omega \,.
\end {align}
Using sum rules it is possible to express $m_1 $, $m_3$ and $m_{-1}$ as \cite{sum_rules_and_giant_resonances_in_nuclei-lipparini_stringari89}
\begin{align}
  m_1 &= \frac12  \av {
    \left[
      \hat F,
      \left[
        \hat H, \hat F
      \right]
    \right]
  }
  \label{eq:m_1}
  \\
  m_3 &= \frac12
  \av{
    \left[
      \left[
        \hat F,\hat H
      \right],
      \left[
        \hat H,
        \left[
          \hat H, \hat F
        \right]        
      \right]
    \right]
  }
  \label{eq:m_3}
  \\
  m_{-1} &= -\frac12\mbox{Im} \;\chi (\hat F, \omega \to 0)
\end{align}
where $\chi(\hat F, \omega)$ is the response function for the operator $\hat F$ at frequency $\omega$.

The SMA corresponds to the assumption that $S (\omega) $ is dominated
by one sharp peak. Then the collective mode excited by the operator
$\hat F $ has  a frequency that is given by $\omega_{\hat F} =
\omega_{p,p-2} = \sqrt{\frac{m_p}{m_{p-2}}}$ for {\it any} value of
$p$. If the SMA is not exact, and more than one peak is present, one
finds that $\omega_{p,p-2} > \omega_{q,q-2}$ for $p> q$ and this approach to finding the collective mode frequency using sum rules becomes less accurate. Nevertheless, $\omega_{p, p-2}$ still provides an upper bound for the lowest frequency collective mode.
We  use both $\omega_{1, -1} $ and $\omega_{3,1} $  to obtain upper
bounds for the collective mode frequencies.

In order to investigate the dipole and the breathing modes
we consider two sets of operators $\hat F $. For the dipole mode we consider $\hat F_d(\theta) = \cos\theta\sum_i x_i + \sin\theta \sum_j x_j $ where indices $i$ run over all the $\ua$ particles and $j$ over all the $\da $ particles. For the breathing modes we consider $\hat F_b (\theta) = \cos\theta\sum_i x_i^2 + \sin\theta \sum_j x_j^2 $.
$\hat F_d$ corresponds to applying linear potential $V_\ua(x) = x \cos \theta$ $, V_\da (x) = x \sin\theta$ and $\hat F_b $ corresponds to applying a quadratic potential $V_\ua(x) = x^2 \cos \theta$ $, V_\da (x) = x^2 \sin\theta$.

Here the single mode approximation amounts to assuming that the
operator $\hat F (\theta) $ has at most two peaks in the response
spectrum, each of which can be made to vanish by suitable choice of
$\theta $. Hence the collective mode frequencies of the two associated
modes can be obtained by maximising or minimising $\omega_{p,
  p-2}$. The lower frequency mode for each $\hat F $ is the density
mode and the higher frequency mode is the spin mode.
\footnote{For the polarized system $p\neq 0$, the modes are mixtures of both density and spin modes. Our terminology refers to the dominant component of the mode for weak interactions where the distinction becomes well defined. }

For $\hat F_d $ we obtain in the continuum limit two values of $\theta
$ which maximise or minimise $\omega_{3, 1} $.  For both $T=0$ and for $T>0$ these are
$\theta_{\text{sd}} = -\arctan\frac{k_{F,\ua}}{k_{F, \da}}$ for the
spin dipole mode and $\theta_{\text{Kohn}} = \frac\pi4$ for the Kohn mode.
For the breathing mode such simple analytical formulae for
$\theta_{sb} $ (spin breathing mode) and $\theta_{db} $ (density
breathing mode) which maximise/minimise $\omega_{p, p-2} $ could not be obtained. These can be obtained numerically though as has been done in Ref. \cite{phase_diag_of_str_int_pol_fg_1d_drummond07}, where the density breathing mode frequency was investigated using sum rules.

We have used our results to address the question: how accurate is the SMA in describing the results of the BdG+RPA approximation?

\begin{figure}[tb]
  \centering
  \includegraphics[width = 4cm]{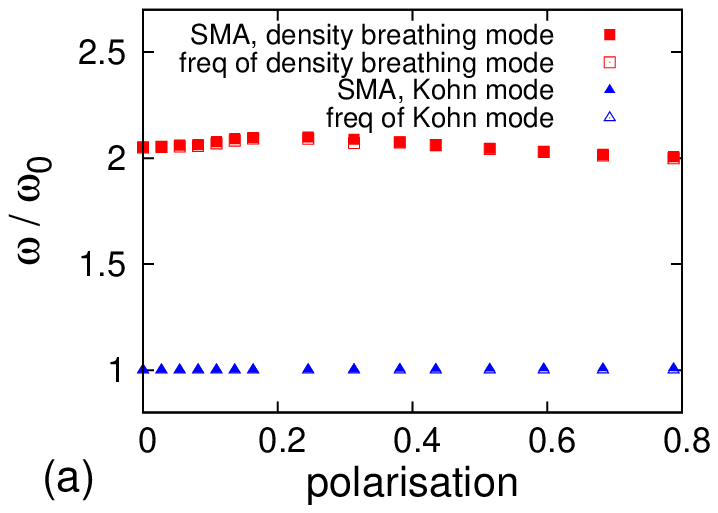}
  \includegraphics[width = 4cm]{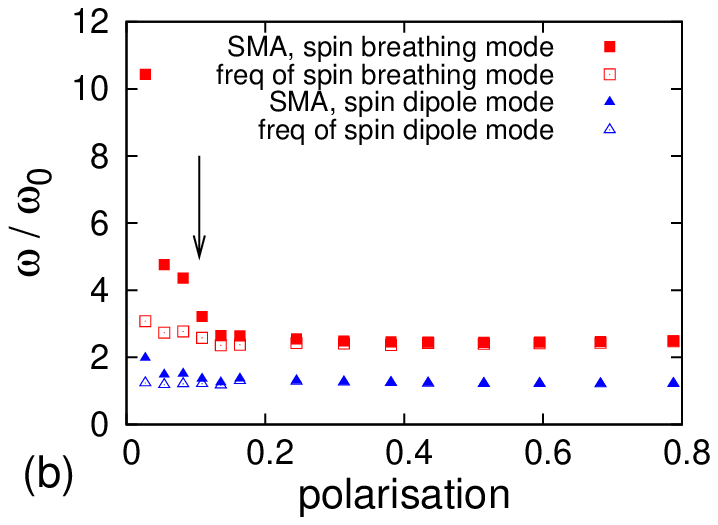}
  \caption{(Colour online)  Calculation of the accuracy of the SMA using $\omega_{1,-1}$. (a): breathing modes and (b): spin modes. These plots are for $\gamma=1.2$ and different polarisations.
    For the density modes (a) the single mode approximation is very accurate.   
 For the spin modes (b) the single mode approximation becomes inaccurate at small polarisations.
    The arrow indicates the point where the outside of the cloud goes from being fully paired ($p\lesssim 0.1$) to being fully polarised ($p\gtrsim 0.1$)}.
  \label{fig:acc_of_sma}
\end{figure}

For all the modes mentioned above  we find that the SMA is strictly speaking not correct,  since $\omega_{3,1}\not=\omega_{1,-1}$.
To gauge the accuracy of the SMA we have compared the prediction of
the sum rule for $w_{1,-1}$ with the  position of the peak in the
response spectrum, see Fig.~\ref{fig:acc_of_sma}. We choose $\omega_{1,-1}$ because it gives higher accuracy results as it is not affected as much by low weight high frequency peaks as $\omega_{3,1}$ is.
For the two density modes which are the lower frequency modes of the two operators $F_d$ and $F_b$ we find that there is very good agreement of the positions of the peak as obtained from the response spectrum and the SMA. For the density dipole mode (the Kohn mode) the sum rule calculation captures the position of the peak up to our numerical accuracy. For the density breathing mode the accuracy typically is about 1\%.  In the limit $p\to 1$ the single mode approximation becomes exact for these two modes as can easily be verified by evaluating eqs.~(\ref{eq:m_1}) and (\ref{eq:m_3}) for a non-interacting gas.

However, for the spin modes the SMA is not as accurate. In particular
for lower polarisations the frequency from the SMA estimate $\omega_{1,-1}$  
can differ quite significantly from the position of the lowest frequency peak  of the full spectrum. This is
due to the presence of many small amplitude high frequency peaks.
The  increased weight of these high frequency peaks and the associated failure of the SMA is correlated with an increase in size of the fully paired region towards the edge of the cloud.

\subsection  {Collective mode signature of the FFLO phase}

\begin{figure}
  \centering
  \includegraphics[width=4.25cm]{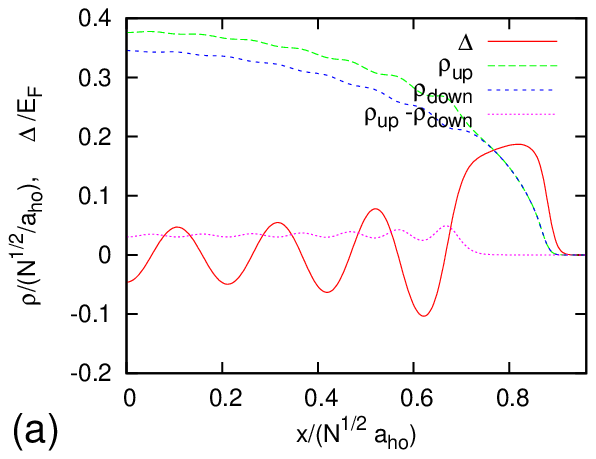}
  \hspace{0.1cm}
  \includegraphics[width=4.cm]{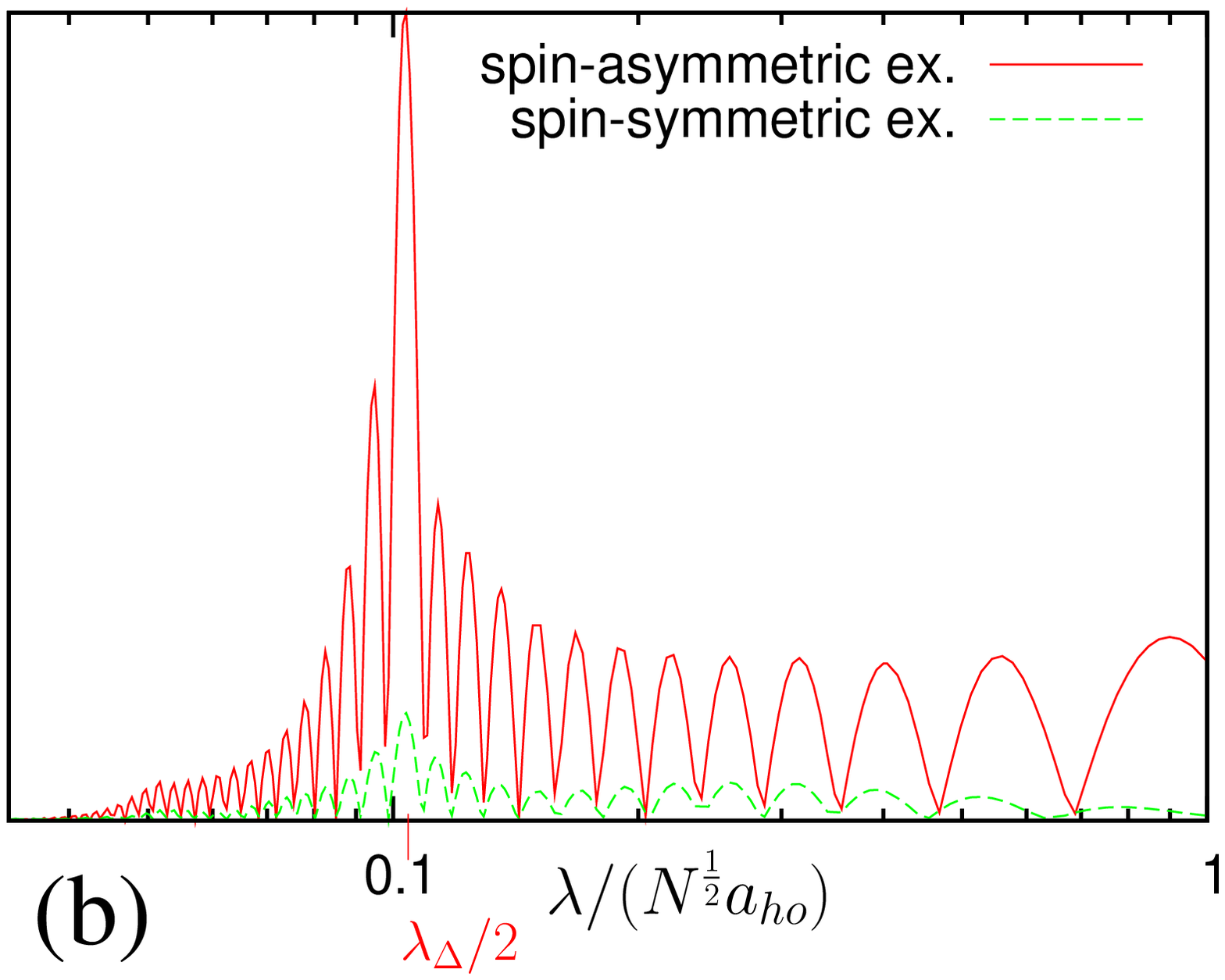}
  \caption{(Colour online.) Configuration with fully paired phase towards the edge of the
    trap. (a): densities and the value of the superfluid gap
    $\Delta$. (b): response of the spin-dipole mode to
    excitations of different wavelengths (arbitrary units). Here $p=0.048$, $\gamma = 0.93$ (measured in centre), number of particles $N = 290$, lattice spacing $a = 3.3\cdot10^{-3} N^{\frac12}a_{ho} $ with $a_{ho} = \sqrt{\frac{\hbar}{m\omega_0}}$. 
    The perturbing potential
    has a fixed amplitude, while the wavelength is varied. }
  \label{fig:full_paired_towards_edge}
\end{figure}

\begin{figure}
  \centering
  \includegraphics[width=4.25cm]{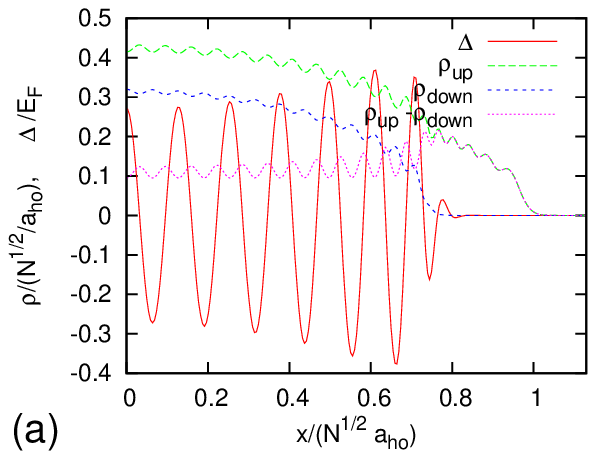}
  \hspace{.1cm}
  \includegraphics[width=4.cm]{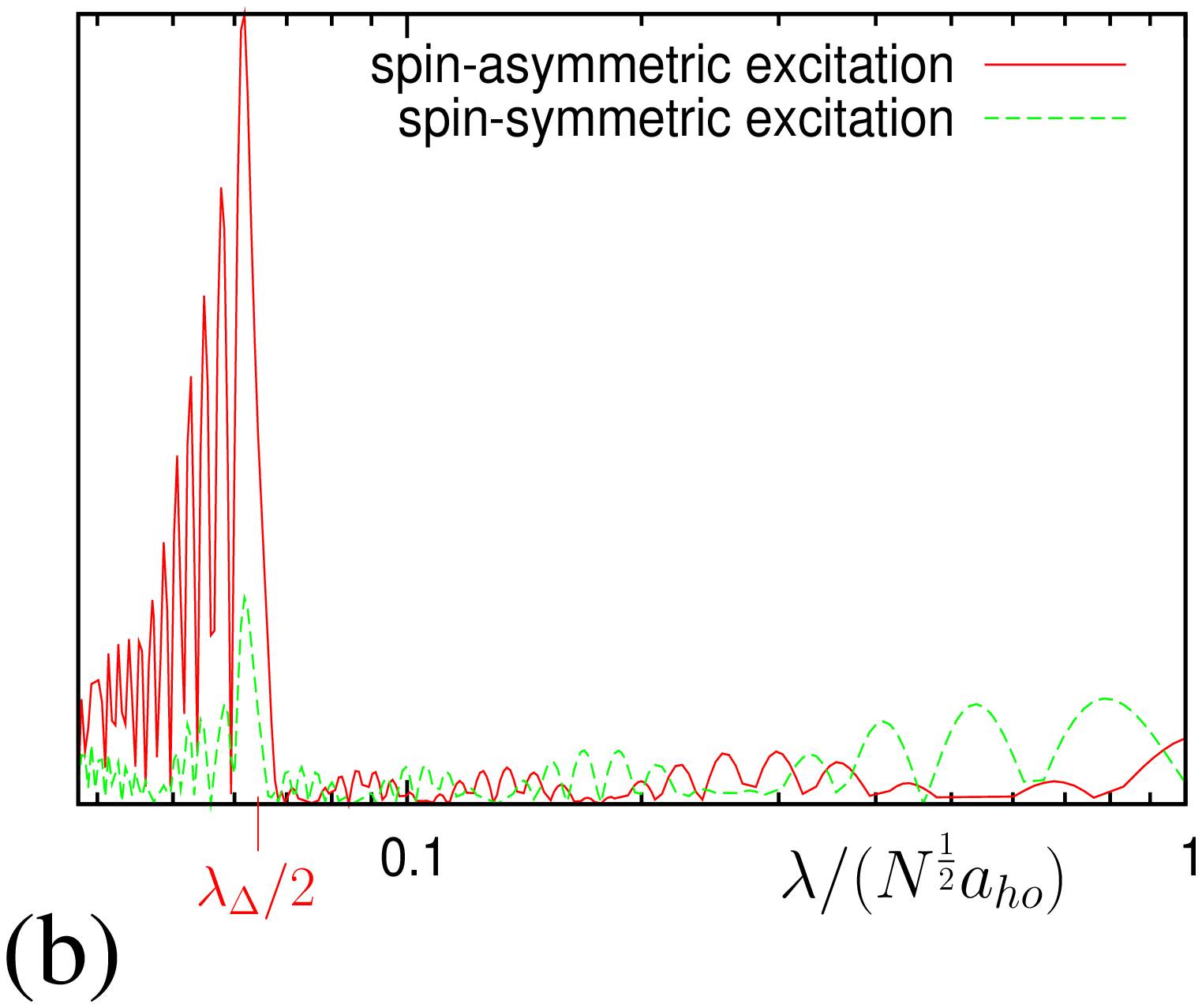}
  \caption{(Colour online.) Configuration with fully polarised phase towards the edge. Here $p=0.25$, $\gamma = 1.5$, $a/(N^{\frac12}a_{ho}) = 4.7\cdot10^{-3}$, $N=143$ . Otherwise the same as Fig.~\ref{fig:full_paired_towards_edge}.}
  \label{fig:full_polarized_towards_edge}
\end{figure}

As described in \cite{signature_of_fflo_first_paper_jonathan_nigel} at
$T = 0 $ we find a dramatic signature of the FFLO phase of a trapped Fermi gas
 when the spin-dipole mode is excited by a short-wavelength potential. When perturbing with
potentials of the form (\ref{eq:vlambda}) we find a large response of
the spin dipole mode when the wavelength of the perturbing potential
becomes $\lambda_\Delta/2$
\footnote{Fig.~\ref{fig:full_polarized_towards_edge}~(b) is based on the same data as Fig.~3~(b) in \cite{signature_of_fflo_first_paper_jonathan_nigel}, however the label for $\lambda_\Delta/2$ is incorrectly placed in \cite{signature_of_fflo_first_paper_jonathan_nigel}. Fig.~\ref{fig:full_polarized_towards_edge}~(b) is correct.}
\footnote{The phase of the perturbation chosen in eq.~(\ref{eq:vlambda}) implies that the perturbation has odd parity about the trap centre so only odd parity modes can be excited. If the phase is chosen such that the perturbation has even parity about the trap centre the low frequency mode excited is the spin breathing mode, but the signature is the same.}.
This response is independent of whether the ground state configuration
is of type (a) or (b), that is whether the outside is fully paired or fully polarised. This response for these two types of system is shown in Figs.~\ref{fig:full_paired_towards_edge} and \ref{fig:full_polarized_towards_edge}.
This response can be
understood in terms of coupling to the gapless modes at
$k=\frac{4\pi}{\lambda_\Delta} = k^*$ of the homogeneous system (see
Fig.\ref{fig:two_sound_modes} and Sec.~\ref{sec:signature-fflo-phase}). A small deviation from $k^*$ (of the
order the inverse system size) allows mixing of this mode to the
spin-dipole mode, causing the response at $k^*$ to be apparent in the
dipolar motion of the atomic cloud.

\begin{figure*}[t]
  \centerline{
    \verysmallpic{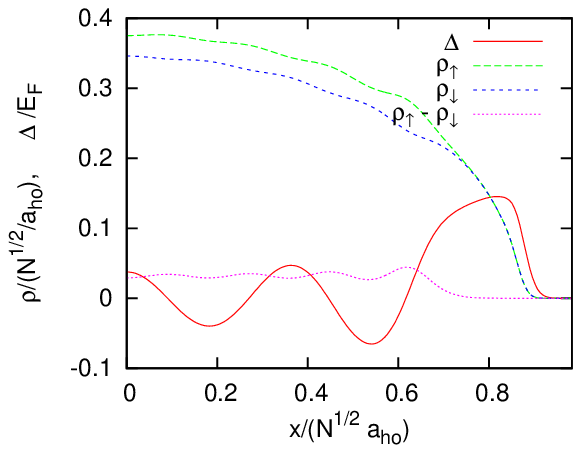}
    \verysmallpic{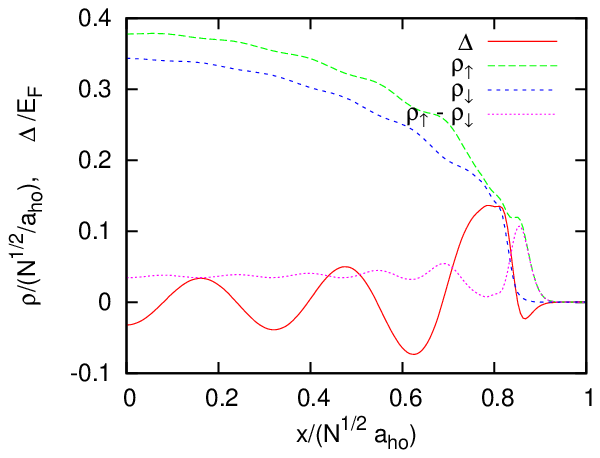}
    \verysmallpic{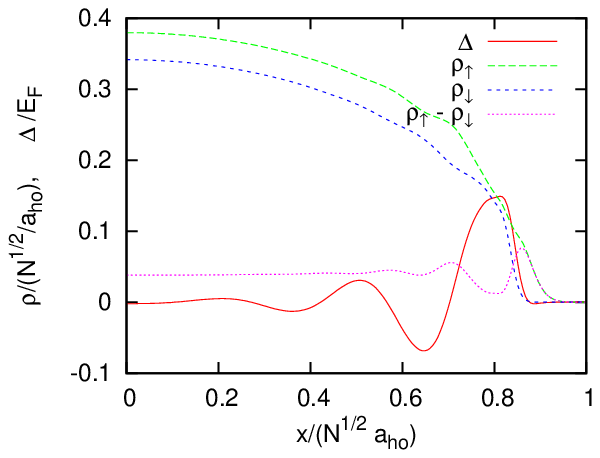}
    \verysmallpic{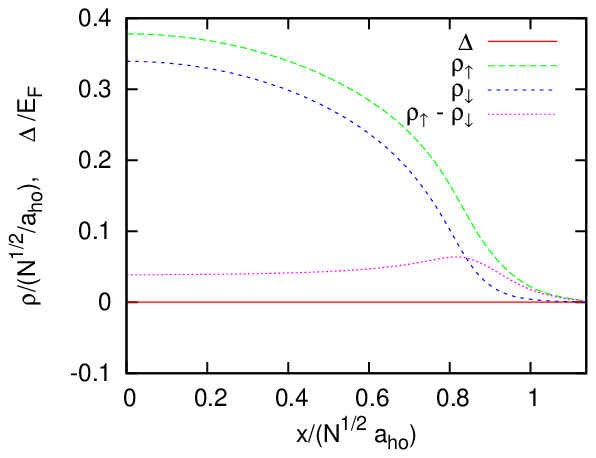}
  }
  \subfigure[]{\verysmallpic{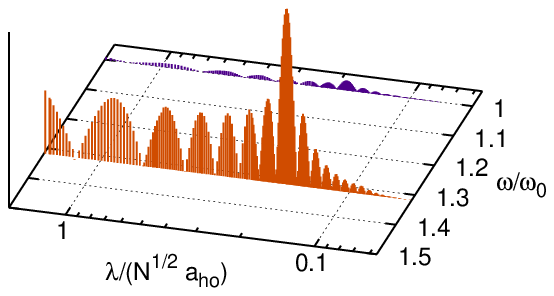}}
  \subfigure[]{\verysmallpic{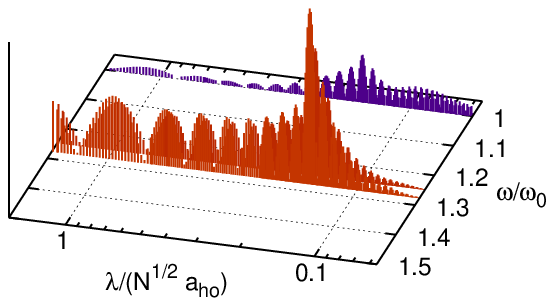}}
  \subfigure[]{\label{fig:resp_trap_intermed_temp}\verysmallpic{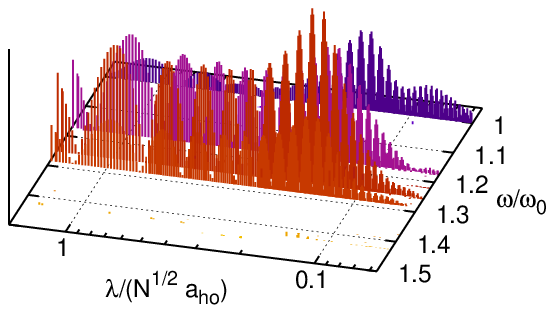}}
  \subfigure[]{\verysmallpic{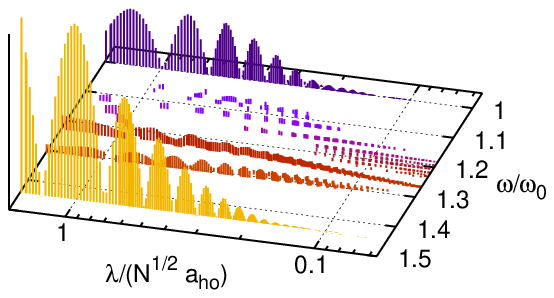}}

  \caption{(Colour online) Response of a system (for which at $T=0$ the outside is fully paired) to a periodic potential of wavelength $\lambda$ and freqency $\omega$. There are 173 particles, $\gamma=0.93$, $p$ varies from 0.05 to 0.09. The temperature is $T = 0.0005T_F$, $T=0.005T_F$, $T=0.02T_F$ and $T=.106T_F$.  For the lowest temperature the spin-dipole frequency is $\omega = 1.31\omega_0$. 
 The sharp peak at the spin-dipole frequency at $\lambda/(N^{1/2} a_{ho}) \approx 0.2$ is the signature of the FFLO phase. At higher temperatures, this sharp response disappears and additional excitations emerge. The case where the outside is fully polarised towards the edge at $T=0$  is qualitatively the same.
  }
  \label{fig:finite_temp_trap}
\end{figure*}

As in the untrapped system, at  very low temperatures the system behaves in the same way as for $T = 0 $.
With increasing temperature the oscillating gap $\Delta$
disappears. Along with it the sharp response at the wavelength
$\lambda_\Delta/2$ disappears, see
Fig.~\ref{fig:finite_temp_trap}. There are a few  additional noteworthy points. As the temperature is increased the number of modes which can
be excited increases. This can lead to the splitting of, for example,
the spin dipole mode into two near degenerate sub modes. For yet
higher temperature further fractionalisation of the modes occurs, see Fig.~\ref{fig:finite_temp_trap}.  At intermediate temperatures (Fig.~\ref{fig:resp_trap_intermed_temp}) an oscillating gap can still be made out  but the peak in the response of the spin dipole mode is strongly attenuated, which may increase the difficulty of finding the FFLO phase.
An example of a system which at $T=0$ is fully paired towards the
outside is given in Fig.~\ref{fig:finite_temp_trap}, but the case
where the outside is fully polarised at $T=0 $ is qualitatively similar.

As in the homogeneous case discussed in Sec.~\ref{sec:signature-fflo-phase} our calculation is performed in the collisionless regime. Using eq.~(\ref{eq_for_T_m_for_collless_regime}) and  the LDA by taking the system size to be roughly the size of the cloud we find that the collisionless regime should be valid up to a temperature of $T/T_F \approx 0.03$. This is slightly lower than in Sec.~\ref{sec:temp-depend} since the system size is larger. Nevertheless,  the three configurations in Fig.~\ref{fig:finite_temp_trap} with an FFLO phase are still  in the collisionless regime.

Most results presented in this paper are for values of $\gamma
  \approx 1 - 1.6$ where the mean-field theory we use is expected to
  be accurate. (For the groundstate, this is confirmed by comparisons
  with exact results \cite{liu_drummond_07}.)  We have performed  
calculations with larger $\gamma $, in the range of $\gamma =2.2 - 3.8$, and
  find qualitatively the same results. While we cannot trust the
  quantitative results of mean-field theory in regimes of strong interactions, particularly at
  unitarity, we believe that the qualitative results remain valid. In
  particular, the key signature of the FFLO phase that we predict
  relies only on the commensurability between the wavelength of the perturbing
  potential and the intrinsic periodicity of the FFLO phase, and is expected to be robust.

\subsection{Experimental considerations}
\label{sec:exper-constr}

Recently experiments have produced true 1D ultracold Fermi gases as an array of 1D tubes \cite{spin_imbalance_in_a_1d_FG-Liao_Hulet_2009}. By changing the
coupling between the tubes it is possible to tune continuously between
a 3D system and a true 1D system.  The experimental results show evidence for the appearance of a partially polarised phase in a regime of parameters consistent with 
the theoretical expectations for the phase associated with the FFLO state.
Probing the collective modes in the manner described in this paper 
would lead to direct evidence of the intrinsic modulation of the FFLO phase.

Observing the signature of the FFLO phase in experiments  in this way requires  the ability to create a variable wavelength
optical lattice. This can be done using the technique used by Ref.
\cite{ex_spectrum_of_bec_steinhauer_davidson02}.  Although the signal
appears for a spin-symmetric perturbation,
the signal is clearer for
spin-asymmetric perturbations. Thus, any spin-dependence of the optical lattice will
enhance the ability to distinguish the peaks associated with the
oscillation of $\Delta$ from the background peaks.  For Bosons ($^{87}$Rb) potentials with spin-dependence
have been created by Ref.
\cite{cohenrent_transp_of_neut_atoms_in_spin_dep_opt_latt_mandel_bloch03},
and would be ideally suited to this purpose.
There are additional complications associated with applying
spin-dependent potentials to $^6$Li due to  increased  loss rates
resulting from the application of the potentials
\cite{bragg_scattering_and_spin_struc_fac_in_2cmpt_gases-carusotto2006}. As
discussed in Ref.  \cite{bragg_scattering_and_spin_struc_fac_in_2cmpt_gases-carusotto2006} it is nevertheless in principle possible to set up these potentials.
There are two natural ways to find the response of the spin-dipole
mode.  One way is to  follow  the approach of the calculation precisely,
and make the strength of the optical lattice time dependent,
as in
Ref. \cite{transitino_from_strongly_int_1d_sf_to_MI_stoferle_Kohl_Esslinger04}.
The spin-dipole mode can then be selectively excited by bringing the
temporal oscillation of the lattice in resonance with the spin-dipole
mode.  Alternatively   a static periodic potential can be applied to the
system and the system  allowed  to equilibrate. Subsequently  this potential is
abruptly switched off (similar to the approach used in Ref.
\cite{altmeyer07}).  This will excite collective modes of many
frequencies from which the response of the spin-dipole mode can be
obtained by Fourier transform.

In \cite{quasi_1d_pol_fermi_SFs-parish-huse07} it is maintained that in order to see the FFLO phase by in situ imaging it is advantageous to induce coupling between the different 1D tubes in order to align the phases of the FFLO state between the different tubes. However, for the method we are considering this should not be necessary. The response of the system depends on the wavelength with which the system is excited and on $\lambda_\Delta$. Only one laser is used to create the optical lattice with which the system is excited which implies that the wavelength with which the system is excited is the same for all tubes. Therefore,  as long as the wavelength of the FFLO phase is approximately the same between different tubes, the response will be be small or large for the same excitation wavelengths. As a result it should be possible just to work with very weakly coupled or uncoupled 1D tubes and still see this effect.

One experimental difficulty in detecting the FFLO phase in systems of
the type studied in Ref. \cite{spin_imbalance_in_a_1d_FG-Liao_Hulet_2009} -- which is shared by all probes of the intrinsic periodicity of the FFLO phase -- is
the inhomogeneity across the many 1D tubes.
Due to the different positions of
the tubes in the atomic trap, different tubes will have different
chemical potentials. In general this will lead to different values of
$\lambda_\Delta$ in different tubes.
Since the large response occurs for perturbations with $\lambda
=\lambda_\Delta /2$ only those tubes which meet the condition on $\lambda_\Delta $ will show a strong response.
If all the tubes can only be imaged as a whole 
it is therefore desirable to work in a regime where $\lambda_\Delta$
varies minimally between different tubes. As pointed out by Ref.
\cite{quasi_1d_pol_fermi_SFs-parish-huse07} 
it is possible to choose a
point in the phase diagram where $\dro {\lambda_\Delta} \mu = \drt
{\lambda_\Delta} \mu =0$. At this point the value of $\lambda_\Delta$
should be the same over a large range of tubes in the centre of the
system.

Our method provides a very elegant way to overcome the effects of inhomogeneity, if in situ imaging can allow individual or a small collection of 1D tubes to be resolved. This will allow one to relax the condition of having any intertube tunneling and being 
at a special point in the phase diagram.
By exciting
the  whole system with  a given wavelength $\lambda $, 
 only those tubes with  $\lambda_\Delta \simeq 2\lambda$ will have
their spin-dipole oscillations excited.
 Tubes with the same value for $\lambda_\Delta$ lie on
concentric cylinders around the trap centre.
Thus, if imaging of the cloud has sufficient resolution to identify which tubes are oscillating, one should 
see response from tubes that lie on a cylindrical region around the axis of the trap. Changing the wavelength $\lambda$ will change the radius of this cylindrical region. In
this way the FFLO wavelength could be studied as a function of the
position of the 1D tube in the trap. This approach is only limited by the power
to resolve the 1D tubes.

\subsection {Modulation of the coupling constant}

We have investigated the response of the one-dimensional Fermi gas to a modulation of the coupling constant. This is an interesting perturbation to consider, since  using Feshbach resonances it  is relatively easy to implement a uniform modulation of the coupling constant experimentally. Recently it was also found, that in a lattice system parametric modulation can be achieved by modulating the lattice depth \cite{parametric_resonance_in_1d_fermionic_sys-graf-mariani}.

Modulations of the coupling constant  couple to density differences. The FFLO state  has a density difference that shows quasi-periodic spatial oscillations, as explained above. Hence it  is interesting to ask if the response of the system  is sensitive to the appearance of the ordering characteristic of the  FFLO phase.

Within linear response the modulation of the coupling constant of the form $U = U_0 + \delta U(t) $ leads to a perturbation of the form
\begin{align}
  \delta W_{i, \sigma}(t) &= \frac{\delta U(t)}{U_0} V^{int}_{i,\sigma}\\
  \delta \Delta_i(t) &= \frac{\delta U(t)}{U_0} \Delta_i
\end{align}
and hence the response to a modulation of the coupling constant is amenable to our RPA calculation.

 We have studied the response of the trapped gas to modulations of the coupling constant, focusing on the regime of low frequencies.
Since a modulation of the coupling constant does not break reflection symmetry about the centre of the trap it can only excite trap modes of even parity. As a result the lowest  frequency modes which can be excited are the breathing modes.

\begin{figure*}[tb]
  \centering
  \centerline{
    \verysmallpic{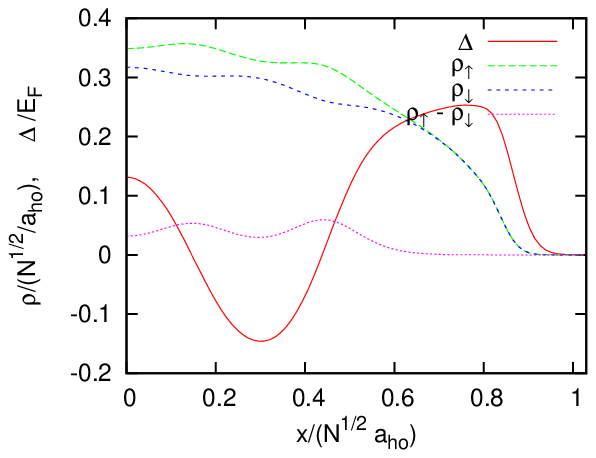}
    \verysmallpic{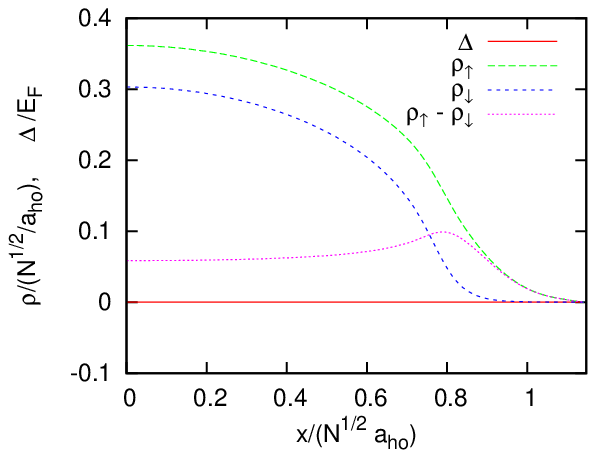}
    \verysmallpic{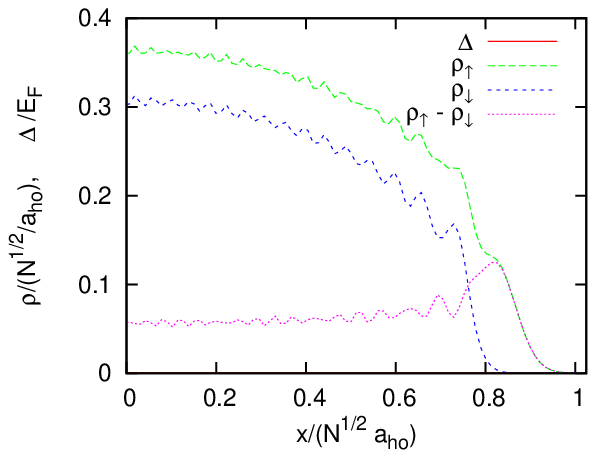}
    \verysmallpic{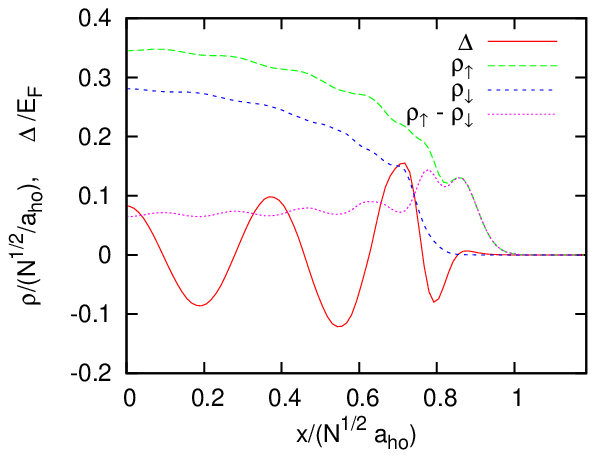}
  }

  \centerline{
    \verysmallpic{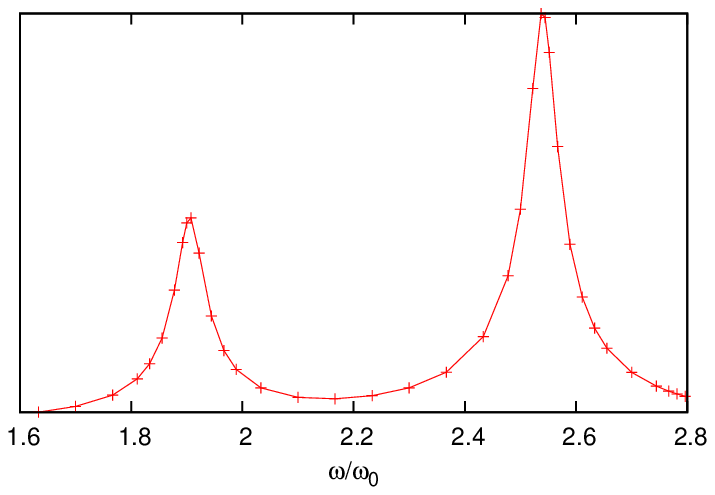}
    \verysmallpic{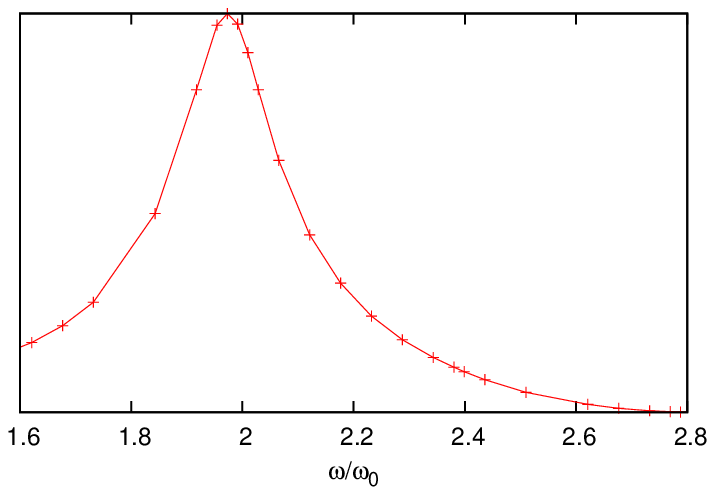}
    \verysmallpic{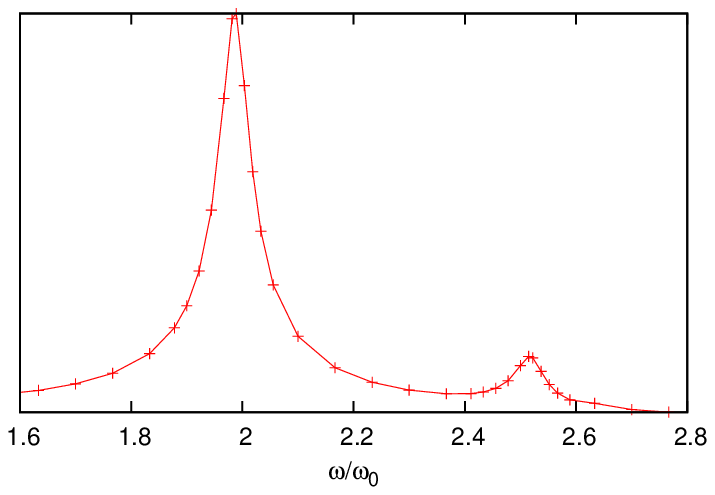}
    \verysmallpic{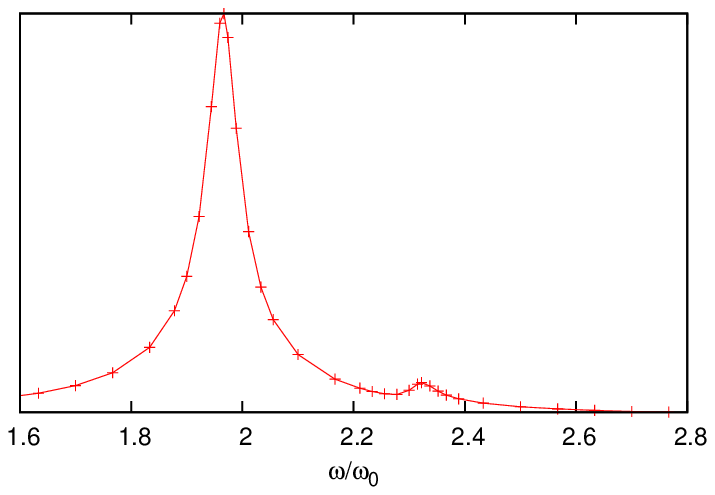}
  }  
  \caption{(Colour online) Response of a trapped system to a modulation of the coupling constant. Whenever the outside is fully paired the weight of the response of the spin breathing mode (the higher frequency mode) is larger than that of the density breathing mode (lower frequency). All are for $\gamma=1.2$. From left to right we have a system at $T = 0 $ which is fully paired towards the edge (73 particles), a high-temperature system ($T=0.1 $ , 74 particles), a system where the order parameter $\Delta$ was set to 0 (74 particles) and a system with FFLO phase but which is fully polarised at the edge (67 particles). The bottom row shows the respective responses.
}
  \label{fig:mod_of_g}
\end{figure*}

We find that the nature of the response of the system depends strongly
on whether the system is in the configuration (a), which is fully paired toward the edge and exhibits an FFLO phase in the centre, or in the configuration (b), 
which is fully polarised towards the edge. When the outside is fully polarised the majority of the response weight is at the lower frequency density breathing mode and there is little or no response at the higher frequency spin breathing mode. On the other hand, in the case where the outside is fully paired and there is an FFLO phase in the centre the situation is reversed. Most of the weight of the responses on the high-frequency spin breathing mode. See Fig.~\ref{fig:mod_of_g}.

While the nature of the response is senstive to the global distribution of particles, whether or not an FFLO phase is present makes little difference to the response spectrum. We must therefore conclude that  there is no clear signal of the presence of the FFLO phase in the collective mode spectrum excited by a modulation of the coupling constant. On the other hand the collective mode spectrum appears to be sensitive to whether the outside region of the cloud is fully paired or fully polarised.

\section {conclusion}

To conclude, we have investigated the collective mode spectrum of the
imbalanced Fermi gas. The FFLO phase can be detected due to the
formation of a characteristic periodicity of the densities.
This makes it possible to use short wavelength perturbations which couple to the intrinsic periodicity of the FFLO phase in order to excite long wavelength responses which are easy to measure in experiment.
This signal in the collective mode spectrum persists up to the temperature where the FFLO phase disappears.
We also investigated the accuracy of the single mode approximation. We found that it works very well for density modes, but not as well for spin modes. Finally, we found that a modulation of the coupling constant results in a collective mode spectrum which allows us to distinguish long wavelength characteristics of the density profile but not short wavelength characteristics, as are associated with the FFLO phase.

\acknowledgments{We would like to thank Carlos Lobo for suggesting to investigate the modulation of the coupling constant and for stimulating discussions. This work was supported by EPSRC Grant No. EP/F032773/1.}


\begin{thebibliography}{10}

\bibitem{bloch07}
I. Bloch, J. Dalibard, and W. Zwerger, Rev. Mod. Phys. {\bf 80},  885  (2008).

\bibitem{pitaevskii_stringari_RMP_09}
S. Giorgini, L.~P. Pitaevskii, and S. Stringari, Rev. Mod. Phys. {\bf 80},
  1215  (2008).

\bibitem{pairing_and_phase_sep_in_pol_fg_partridge_hulet}
G. Partridge {\it et~al.}, Science {\bf 311},  503  (2006).

\bibitem{fermionic_sf_with_imbalanced_populations-zwierlein-ketterle06}
M. Zwierlein, A. Schirotzek, C. Schunck, and W. Ketterle, Science {\bf 311},
  492  (2006).

\bibitem{larkin_ovchinnikov_64}
A. Larkin and Y. Ovchinnikov, Zh. Eksp. Teor. fiz. {\bf 47},  1136  (1964),
  [{S}ov. Phys. JETP 20, 762 (1965)].

\bibitem{superconductivity_in_spin_exch_field_fulde_ferrell64}
P. Fulde and R.~A. Ferrell, Phys. Rev. {\bf 135},  A550  (1964).

\bibitem{landau_lifshitz_pitaevskii_stat_phys_part2}
E. Lifshitz and L. Pitaevskii, {\em Statistical Physics Part 2} (Pergamon
  Press, Oxford, 1980).

\bibitem{bec-bcs_xover_in_pol_res_SF-radzihovsky}
D.~E. Sheehy and L. Radzihovsky, Annals of Physics {\bf 322},  1790   (2007).

\bibitem{orso_attr_fermi_gases_bethe_ansatz07}
G. Orso, Phys. Rev. Lett. {\bf 98},  070402  (2007).

\bibitem{phase_diag_of_str_int_pol_fg_1d_drummond07}
H. Hu, X.-J. Liu, and P.~D. Drummond, Phys. Rev. Lett. {\bf 98},  070403
  (2007).

\bibitem{pairing_states_of_pol_fg_feiguin_meisner07}
A.~E. Feiguin and F. Heidrich-Meisner, Phys. Rev. B {\bf 76},  220508(R)
  (2007).

\bibitem{quasi_1d_pol_fermi_SFs-parish-huse07}
M.~M. Parish, S.~K. Baur, E.~J. Mueller, and D.~A. Huse, Phys. Rev. Lett. {\bf
  99},  250403  (2007).

\bibitem{batrouni_scalettar_09}
G.~G. Batrouni, M.~H. Huntley, V.~G. Rousseau, and R.~T. Scalettar, Phys. Rev.
  Lett. {\bf 100},  116405  (2008).

\bibitem{fflo_state_in_1d_attr_hubbard_model_luescher_laeuchli08}
A. L\"{u}scher, R.~M. Noack, and A.~M. L\"{a}uchli, Phys. Rev. A {\bf 78},
  013637  (2008).

\bibitem{spin_imbalance_in_a_1d_FG-Liao_Hulet_2009}
Y. Liao {\it et~al.}, Spin-Imbalance in a One-Dimensional Fermi Gas,
  arXiv:0912.0092v1, 2009.

\bibitem{spectral_signatures_of_fflo_in_1d_bakhtiari_torma}
M.~R. Bakhtiari, M.~J. Leskinen, and P. T\"{o}rm\"{a}, Phys. Rev. Lett. {\bf
  101},  120404  (2008).

\bibitem{signature_of_fflo_first_paper_jonathan_nigel}
J.~M. Edge and N.~R. Cooper, Physical Review Letters {\bf 103},  065301
  (2009).

\bibitem{normal_state_of_a_polarized_FG_at_unit_Lobo06}
C. Lobo, A. Recati, S. Giorgini, and S. Stringari, Phys. Rev. Lett. {\bf 97},
  200403  (2006).

\bibitem{collective_osc_of_imb_FG_polaron_nascimbene_salomon}
S. Nascimb\`ene {\it et~al.}, Phys. Rev. Lett. {\bf 103},  170402  (2009).

\bibitem{coll_ex_of_trapped_imb_FG-schaeybroeck08}
A. Lazarides and B. Van~Schaeybroeck, Phys. Rev. A {\bf 77},  041602(R)
  (2008).

\bibitem{trapped_2d_FG_with_pop_imbalance-schaeybroeck09}
B. {Van Schaeybroeck}, J. Tempere, and A. Lazarides, Trapped Two-Dimensional
  Fermi Gases with Population Imbalance, arXiv:0911.0984v1, 2009.

\bibitem{Phase_trans_pairing_sig_attractive_Fermi_gases-guan07}
X.~W. Guan, M.~T. Batchelor, C. Lee, and M. Bortz, Phys. Rev. B {\bf 76},
  085120  (2007).

\bibitem{exact_anal_of_delta_fn_spin_half_attr_FG-Iida08}
T. Iida and M. Wadati, Journal of the Physical Society of Japan {\bf 77},
  024006  (2008).

\bibitem{magnetism_and_quantum_phase_trans-he_batchelor09}
J.-S. He, A. Foerster, X.~W. Guan, and M.~T. Batchelor, New Journal of Physics
  {\bf 11},  073009 (17pp)  (2009).

\bibitem{anal_thermodyn_and_thermometr_of_gaudin_yang_gases-zhao_oshikawa09}
E. Zhao {\it et~al.}, Phys. Rev. Lett. {\bf 103},  140404  (2009).

\bibitem{some_exact_results_for_many_body_problem_in_1d_w_delta_int-yang_prl67}
C.~N. Yang, Phys. Rev. Lett. {\bf 19},  1312  (1967).

\bibitem{un_systeme_a_une_dimension_de_fermions_en_interaction-gaudin67}
M. Gaudin, Phys. Lett. A {\bf 24},  55  (1967).

\bibitem{thermodyn_of_1d_solvable_models-takahashi}
M. Takahashi, {\em Thermodynamics of One-Dimensional Solvable Models}
  (Cambridge University Press, Cambridge, UK, 1999).

\bibitem{liu_drummond_07}
X.-J. Liu, H. Hu, and P.~D. Drummond, Phys. Rev. A {\bf 76},  043605  (2007).

\bibitem{bruun_mottelson01}
G.~M. Bruun and B.~R. Mottelson, Phys. Rev. Lett. {\bf 87},  270403  (2001).

\bibitem{Yang_inhomogeneous_sc_state_1d_01}
K. Yang, Phys. Rev. B {\bf 63},  140511(R)  (2001).

\bibitem{mizushima07}
T. Mizushima, M. Ichioka, and K. Machida, Journal of the Physical Society of
  Japan {\bf 76},  104006  (2007).

\bibitem{finite_temp_phase_diag_of_spin_pol_fg-liu_drummond08}
X.-J. Liu, H. Hu, and P.~D. Drummond, Phys. Rev. A {\bf 78},  023601  (2008).

\bibitem{sound_velocity_and_dim_crossover_in_sf_FG-koponen-Torma}
T. Koponen, J.-P. Martikainen, J. Kinnunen, and P. T\"{o}rm\"{a}, Physical
  Review A (Atomic, Molecular, and Optical Physics) {\bf 73},  033620  (2006).

\bibitem{forster75}
D. Forster, {\em Hydrodynamic Fluctuations, Broken Symmetry, and Correlation
  Functions} (W. A. Benjamin, Reading, Mass., 1975).

\bibitem{fflo_pairing_in_1d_opt_latt-rizzi_fazio08}
M. Rizzi {\it et~al.}, Phys. Rev. B {\bf 77},  245105  (2008).

\bibitem{HaldaneJPhysC}
F.~D.~M. Haldane, Journal of Physics C: Solid State Physics {\bf 14},  2585
  (1981).

\bibitem{theory_of_quantum_liquids1_nozieres_1966}
P. Nozi\`{e}res and D. Pines, {\em The theory of quantum liquids} (Perseus
  Books, New York, 1966).

\bibitem{quantum_physics_in_1d_giamarchi}
T. Giamarchi, {\em Quantum Physics in One Dimension} (Oxford University Press,
  Oxford, 2004).

\bibitem{coll_props_of_pol_FG-bruun-stringari08}
G.~M. Bruun {\it et~al.}, Phys. Rev. Lett. {\bf 100},  240406  (2008).

\bibitem{sum_rules_and_giant_resonances_in_nuclei-lipparini_stringari89}
E. Lipparini and S. Stringari, Physics Reports {\bf 175},  103   (1989).

\bibitem{ex_spectrum_of_bec_steinhauer_davidson02}
J. Steinhauer, R. Ozeri, N. Katz, and N. Davidson, Phys. Rev. Lett. {\bf 88},
  120407  (2002).

\bibitem{cohenrent_transp_of_neut_atoms_in_spin_dep_opt_latt_mandel_bloch03}
O. Mandel {\it et~al.}, Phys. Rev. Lett. {\bf 91},  010407  (2003).

\bibitem{bragg_scattering_and_spin_struc_fac_in_2cmpt_gases-carusotto2006}
I. Carusotto, Journal of Physics B: Atomic, Molecular and Optical Physics {\bf
  39},  S211  (2006).

\bibitem{transitino_from_strongly_int_1d_sf_to_MI_stoferle_Kohl_Esslinger04}
T. St\"oferle {\it et~al.}, Phys. Rev. Lett. {\bf 92},  130403  (2004).

\bibitem{altmeyer07}
A. Altmeyer {\it et~al.}, Phys. Rev. A {\bf 76},  033610  (2007).

\bibitem{parametric_resonance_in_1d_fermionic_sys-graf-mariani}
C.~D. Graf, G. Weick, and E. Mariani, Parametric resonance and spin-charge
  separation in 1D fermionic systems, arXiv:0910.4123v2, 2009.

\end{thebibliography}
\end{document}